\documentclass[aps,pre,superscriptaddress,twocolumn,showpacs,nofootinbib]{revtex4}

\usepackage{bm}%
\usepackage[colorlinks=true,linkcolor=red]{hyperref}
\usepackage{amsmath}
\usepackage{mathtools}
\usepackage{amsfonts}
\usepackage{amssymb}
\usepackage{verbatim}
\usepackage[english]{babel}
\usepackage{multirow}
\usepackage{epsfig}
\usepackage{graphicx}
\usepackage{subfig}
\usepackage[sort&compress]{natbib}
\usepackage{times}

\newcommand{\newc}{\newcommand}
\newc{\beq}{\begin{equation}}
\newc{\eeq}{\end{equation}}
\newc{\kt}{\rangle}   
\newc{\br}{\langle}
\newc{\ld}{\lambda}

\begin{document}

\title{\bf\noindent Random pure states: quantifying bipartite entanglement beyond the linear statistics}
\author{Pierpaolo Vivo}
\affiliation{Department of Mathematics, 
King's College London,
Strand,
London
WC2R 2LS, UK}
\email{pierpaolo.vivo@kcl.ac.uk}
\author{ Mauricio P. Pato}
\affiliation{Inst\'ituto de F\'isica, Universidade de S\~ao Paulo
Caixa Postal 66318, 05314-970 S\~ao Paulo, S.P., Brazil}
\email{mpato@if.usp.br}
\author{Gleb Oshanin}
\affiliation{Sorbonne Universit\'es, UPMC Univ Paris 06, UMR 7600, LPTMC, F-75005, Paris, France}
\affiliation{CNRS, UMR 7600, Laboratoire de Physique Th\'{e}orique de la Mati\`{e}re Condens\'{e}e, F-75005, Paris, France}
\email{oshanin@lptmc.jussieu.fr}

\date{\today}

\begin{abstract}
We analyze  the properties of entangled random pure states
of a quantum system partitioned into two smaller subsystems of dimensions $N$ and $M$. 
Framing the problem in terms of random matrices with a fixed-trace constraint,  
we establish, for arbitrary $N \leq M$,
a general relation between the $n$-point densities and the cross-moments 
of the eigenvalues of the reduced density matrix, i.e. the so-called Schmidt eigenvalues, and 
the analogous functionals of the  eigenvalues
of the Wishart-Laguerre ensemble of the random matrix theory. 
This allows us to
derive explicit expressions for two-level densities, 
and also an exact expression for the variance of von Neumann entropy at finite $N,M$.
Then we focus on the moments $\mathbb{E}\{K^a\}$ of the Schmidt number $K$, the reciprocal of the purity. This is a random variable supported on $[1,N]$, which quantifies
the number of degrees of freedom effectively contributing to the entanglement. We derive a wealth of analytical results for $\mathbb{E}\{K^a\}$ for 
$N = 2$ and $N=3$ and arbitrary $M$, and also for square $N = M$ systems
 by spotting for the latter a connection with the probability $P(x_{min}^{GUE} \geq \sqrt{2N}\xi)$ that the smallest eigenvalue $x_{min}^{GUE}$ of a $N\times N$ matrix belonging to the Gaussian Unitary Ensemble is larger than $\sqrt{2N}\xi$. As a byproduct, we present an exact asymptotic expansion for $P(x_{min}^{GUE} \geq \sqrt{2N}\xi)$ for finite $N$ as $\xi \to \infty$. Our results are corroborated by numerical simulations whenever possible, with excellent agreement.

\end{abstract}


\maketitle


\section{Introduction}

Entanglement is perhaps one of the most baffling features of quantum systems. 
Indeed, the possibility of producing entangled states was first 
considered as the signature of the incompleteness of quantum mechanics
~\cite{gen}. 
However, eventually entanglement was verified experimentally and recognized as a valid and fundamental feature of the quantum world.
Moreover,
practical implications of quantum entanglement are foreseen nowadays, e.g., in the fast developing fields of quantum
 information and computation~\cite{NC,CC}. 
There, in order to achieve the highest 
 computational power, it is desirable (at least theoretically) to produce states with large entanglement. 
 
For a bipartite system 
consisting of two subsystems of dimensions $N$ and $M$ (with, e.g., $N \leq M$),
several proxies were introduced to 
quantify
the degree of entanglement,  which are all
functionals of  $N$ non-negative 
eigenvalues $\lambda_i$ of the reduced density matrix, satisfying the normalization 
constraint $\sum_{i=1}^N \lambda_{i}  = 1 $ and called \emph{Schmidt eigenvalues} (see Section \ref{model} for details).  

These functionals are, to name a few, 
the entanglement entropies -
von
Neumann entropy 
\begin{equation}
\label{svn}
S_{\rm vN} =  - \sum_{i=1}^N \lambda_i \ln \lambda_i \,,
\end{equation}
or $q$-parametrized 
R\'enyi entropy\footnote{Note that in the limiting cases when $q \to 1$
or $q \to \infty$ the R\'enyi entropy converges respectively 
to the von Neumann entropy $S_{\rm vN}$ or to $\ln(1/\lambda_{\rm max})$, where $\lambda_{\rm max}$ is the largest eigenvalue.}, 
$S_q = (\ln \Sigma_q)/(1-q)$, where $\Sigma_q = \sum_{i=1}^N \lambda_i^q$; the purity $\Sigma_2$, 
and also its reciprocal value - the so-called Schmidt number
\begin{equation}
\label{K}
K = \dfrac{1}{\sum_{i=1}^N \lambda^2_i } \,,
\end{equation}
defined as the effective number of non-zero coefficients in the Schmidt decomposition (see below), i.e., 
the number of effective degrees of freedom contributing to the entanglement. The Schmidt number is defined on the interval $[1,N]$ and $K \sim N$ corresponds to maximal entanglement.
In some instances,  the Schmidt number can be directly measured experimentally \cite{exter}. 

\emph{Random pure} states, for which 
the Schmidt eigenvalues are strongly correlated random variables,
have attracted a strong interest in recent years:  
they are believed to constitute a promising candidate for quantum 
computation 
since
their \emph{average} entanglement entropy is
 close to the maximal possible entropy of a completely degenerate state, 
 when all $\lambda_i = 1/N$ and hence, $S_{\rm vN} = \ln(N)$~\cite{Lubkin,Page,theo1}. 
 Furthermore, 
 they may serve as  a reference point whose entanglement content is to 
 be compared to an arbitrary quantum state evolving in time. They 
  also appear in the study of quantum chaotic 
 or non-integrable systems~\cite{chaos,BL,penson,theo3}. 
Finally, as we proceed to show, the Schmidt eigenvalues for random pure states 
have the same joint probability density function (jpdf) as the so-called \emph{scaled eigenvalues} (see, e.g., \cite{krish}), the eigenvalues of Wishart random matrices normalized by the trace. The latter
have diverse applications both in statistics and in performance analysis of wireless communication systems, the spectrum sensing problem in cognitive radio networks being just one particular example (for which $N$ and $M$ are the number of antennas and the number of samples per antenna, respectively) \cite{wei2,nadler}. For other results and applications
of entangled random systems, see \cite{theo6,pascazio,FacchiMany,depasquale,cundenflorio}.

As the Schmidt eigenvalues are random variables for random pure states, so are 
the entropies, the purity and the Schmidt number. 
Statistical properties of
the entropies and of the purity, and some other related 
observables (apart from the Schmidt number, which did not receive much attention), have been rather extensively studied
in recent years~\cite{Lubkin,Page,ZS,scott,Giraud1,Giraud2,Znidaric,MBL,Facchi1,Facchi2},
focusing either on the limiting behavior 
when $N$ and $M$ are small or, conversely, 
tend to infinity. 
In particular, the full distribution of the purity has been analyzed  for small $N$ in \cite{Giraud1,Giraud2},
and for large $N$ its characteristic function has been determined 
in \cite{Facchi2}. For square $N = M$ systems with $N \to \infty$, the leading asymptotic behavior of
the distributions of von Neumann and R\'enyi entropies (including large deviation tails) 
was studied in \cite{majlarge,majlarge2} using a Coulomb gas technique. It was there realized that, quite surprisingly, 
even though the average entropy of the
random pure state is close to its maximal value $\ln(N)$, the probability
of this {\em closeness} may be very small. We will comment on this result further on.

For arbitrary, not necessarily large $N$ and $M$, which is often the most relevant case in practice, 
the only available results so far concern 
the spectral  (or one-point) densities \cite{theo3,theo2,theo5,theo4} and moments of the purity \cite{scott,Giraud1,Giraud2}.
Note, however, that the observations  made in
  \cite{majlarge,majlarge2} 
  warn us that \emph{average} values may not be representative of the actual behavior: to gain a better understanding of how meaningful they are, 
 one has to 
 go beyond the linear statistics and estimate the effective broadness of the corresponding distributions. 
This would require, e.g., the knowledge of the variances of these entanglement quantifiers for any $N$ and $M$, 
 which is lacking at present. Moreover, we stress that the von Neumann entropy may not be a 
 proper measure of the degree of entanglement as it exhibits a logarithmic growth with $N$: in the limit $N \to \infty$, it may not be possible to distinguish whether the system attains complete or partial entanglement. The Schmidt number, in particular, seems to be a better quantifier of the degree of entanglement since it grows with $N$ as a power law (see below).

In this paper, we 
focus on non-linear statistics for a bipartite entanglement of random pure states.
We consider first the $n$-point densities\footnote{The notation $[\bm v]_n$ stand for the first $n$ components of the vector $\bm v$.} ${\rho}_n^{(FT)}([\bm\lambda]_n)$ of the Schmidt eigenvalues of arbitrary oder $n$. 
We determine such densities 
using three complementary approaches: the first is    
the  generalization of the method developed previously in \cite{theo4,theo5}
for calculation of the one-point densities of the Schmidt eigenvalues. This allows us to show in a very compact way that   the
$n$-point densities ${\rho}_n^{(FT)}([\bm\lambda]_n)$
can be expressed as a suitable Laplace transform of corresponding $n$-point densities ${\rho}_n^{(WL)}([{\bm y}]_n)$ of
the standard $\beta$-Wishart-Laguerre ($\beta$-WL) ensembles of random matrices.

Our second approach hinges on the (so far seemingly unnoticed) fact 
that the fixed-trace (FT) Schmidt eigenvalues have the same jpdf as the so-called scaled variables (see, e.g., \cite{krish,wei2,nadler}), which also
allows for a very straightforward derivation of the general relation between ${\rho}_n^{(FT)}([\bm\lambda]_n)$ and ${\rho}_n^{(WL)}([{\bm y}]_n)$.

Finally, we establish a link between the cross-moments of the Schmidt eigenvalues of arbitrary order 
and an analogous functional of the scaled variables, which allows us to 
relate the cross-moments of the $\beta$-FT and of the standard $\beta$-WL ensembles. 

Employing these tools, we present 
explicit expressions for the two-point densities and for the variance of the von Neumann entropy. Further on, we focus on the moments $\mathbb{E}\{K^a\}$ of the Schmidt number $K$. We derive exact results 
for systems with $N=2$ and $N = 3$ and arbitrary $M$.  We show that
in systems with a fixed $N$ and $M \to \infty$, the moment of order $a$ tends to $N^{a}$, which implies that the Schmidt number attains its maximal value $N$: this is a signature that such systems become completely entangled in this limit. 

Next, concentrating of square systems with $N = M$, we spot a previously unnoticed connection between $\mathbb{E}\{K^a\}$ and the probability $P(x_{min}^{GUE} \geq \sqrt{2N}\xi)$ that the smallest eigenvalue $x_{min}^{GUE}$ of a $N\times N$ matrix belonging to the Gaussian Unitary Ensemble (see, e.g. \cite{mehta,widom} for more details)  is larger than $\sqrt{2N}\xi$. We show that  $P(x_{min}^{GUE} \geq \sqrt{2 N} \xi)$ with $\xi \to \infty$ can be interpreted as the moment generating function of the purity, while $P(x_{min}^{GUE} \geq \sqrt{2 N} \xi)$ with $\xi \to 0$ is the generating function of the moments of $K$ of order $N^2/2 + m$. The moments of order $a$ lower than $N^2/2$ are determined exactly as a certain integral of $P(x_{min}^{GUE} \geq \sqrt{2 N} \xi)$.  As a byproduct of our analysis here, we present an exact asymptotic expansion for $P(x_{min}^{GUE} \geq \sqrt{2 N} \xi)$ with $\xi \to \infty$ and arbitrary $N$. Finally, capitalizing on the results in \cite{DM} for the large deviation form of $P(x_{min}^{GUE} \geq \sqrt{2 N} \xi)$, we establish the leading asymptotic behavior of the moments of $K$ in the limit $N \to \infty$. We show that in the square systems $\mathbb{E}\{K^a\} \sim (N/2)^{a}$, which signifies that here $K$ attains only half of its maximal value and the square systems are far of being completely entangled. This may apparently explain the paradoxical behavior observed in  \cite{majlarge,majlarge2}.

This paper is organized as follows: In Sec. \ref{model} we describe the random pure state setting and introduce some basic definitions. In Sec. \ref{densities} we focus on the $n$-point densities of the $\beta$-FT ensemble and establish a general relation with the analogous quantities of the $\beta$-WL ensembles. 
Eventually we also derive a series of useful relations between the cross-moments of the FT and $\beta$-WL ensembles.
Next, in Sec. \ref{spec} we present explicit, closed-form expressions for the two-point densities and in Sec. \ref{var} we derive an exact expression for the variance of the von Neumann entropy, valid for any $N$ and $M$.  Sec. \ref{Schmidt} is devoted to the analysis of the Schmidt number $K$. Capitalizing on the known results for the distribution function 
of the purity \cite{Giraud1,Giraud2,majlarge,majlarge2},  
we first present the probability density function (pdf) of $K$ for $N = 2$ and $N=3$ and arbitrary $M$, and also discuss the forms of the right and left  tails of this pdf for square $N = M$ systems in the limit $N \to \infty$. 
From these results, we derive exact expression for the moments $\mathbb{E}\{K^a\}$ of $K$ of arbitrary order for $N = 2$ and $N = 3$ and arbitrary $M$, and analyze their asymptotic large-$M$ behavior. 
Next, taking advantage 
of the established relation between the cross-moments of the FT and WL ensembles,  we find an exact representation of $\mathbb{E}\{K^a\}$ of arbitrary, not necessarily integer order $a$ in $N \times N$ systems via the probability $P(x_{min}^{GUE} \geq \sqrt{2 N} \xi)$ that the smallest eigenvalue in the Gaussian Unitary Ensemble 
is larger than $\sqrt{2 N} \xi$. Lastly, we discuss the asymptotic, large-$N$ behavior of these moments.  In Sec. \ref{conc} we conclude with a brief summary of our results. Some technical results are then confined to the Appendices.

\section{The model and definitions}
\label{model}

A precise definition of our settings is as follows. Consider a quantum state
\begin{equation}
\label{decomp}
|\psi\kt = \sum_{i=1}^N \sum_{j=1}^M x_{i,j} |i^{A}\kt \otimes |j^{B}\kt\ ,
\end{equation}
of a composite system, living in a Hilbert space ${\cal H}_{A \otimes B}^{(N+M)}$, which is bipartite into two smaller Hilbert spaces ${\cal
  H}^{(N)}_A$ and ${\cal H}^{(M)}_B$ of dimensions $N$ and $M$ ($N\leq M$). One example of this setting may be a spin set (the subsystem $A$) in contact with a heat bath (the subsystem $B$).

$\{| i^{A}\kt\}$ and $\{| j^{B}\kt\}$ in \eqref{decomp} are assumed to be two complete bases of ${\cal
  H}^{(N)}_A$ and ${\cal H}^{(M)}_B$, respectively. Therefore, the expansion of $|\psi\kt$ on the direct product of these bases involves coefficients $x_{i,j}$, which are the (in general complex) entries of a rectangular $N \times M$ matrix $X$. 

If $X$ is now promoted to a \emph{random} matrix, the class of \emph{random} states $|\psi\kt$ can be further refined by requiring that

\begin{itemize}
\item $|\psi\kt$ must not be expressible as a direct product of two states belonging to the two subsystems $A$ and $B$ (this condition ensures that $|\psi\kt$ is generically \emph{entangled})\ ,
\item the density matrix of the composite system is simply given by $\rho = |\psi\kt \br \psi|$ with the constraint ${\rm Tr}[\rho] = 1$, or equivalently, $\br \psi|\psi \kt=1$. This condition ensures that $|\psi\kt$ is a \emph{pure} state.
\end{itemize}

The density matrix $\rho$ for an entangled pure state $|\psi\kt$ of a bipartite system can be formally written as
\begin{equation}
\label{purestate}
\rho = \sum_{i,i'=1}^N \sum_{j,j'=1}^M x_{i,j} x^*_{i',j'} |i^{A}\kt \br i'^{A}| \otimes |j^{B}\kt \br j'^{B}| \,.
\end{equation}

It is often convenient to consider the \emph{reduced} density matrix $\rho_A = {\rm Tr}_B[\rho]$ as
\begin{equation}
\rho_A =  {\rm Tr}_B[\rho] = \sum_{j=1}^M \br j^{B}| \rho | j^{B}\kt\ ,
\end{equation}
whose role is to separate the contribution of the subsystem $A$ from the environment $B$. Expectation values of quantum observables involving the subsystem $A$ alone can often be more easily computed invoking $\rho_A$. 

Using now the expression in \eqref{purestate}, one gets
\begin{equation}
\rho_A = \sum_{i,i'=1}^N \sum_{j=1}^M x_{i,j} x^*_{i',j} |i^{A}\kt \br i'^{A}| = \sum_{i,i'=1}^N W_{i,i'} |i^{A}\kt \br i'^{A}| \,,
\end{equation}
where $W_{i,i'}$ are the entries of the $N \times N$ covariance matrix $W = X X^\dagger$, with $^\dagger$ denoting hermitian conjugation. Similarly, one might have obtained the reduced density matrix $\rho_B = {\rm Tr}_A[\rho]$ of the environment in terms of the $M \times M$ matrix $W' = X^\dagger X$. It is easy to prove that $W$ and $W'$ share the same set of $N\leq M$ nonzero eigenvalues $\{\ld_1,\ld_2,\ldots,\ld_N\}$. They are all real and positive , and are called \emph{Schmidt eigenvalues}.  

The \emph{Schmidt decomposition} (SD) then takes the form

\begin{equation}
\rho_A = \sum_{i=1}^N \ld_i |\ld_i^{A}\kt  \br \ld_i^{A}| \ ,
\end{equation} 
where $|\ld_i^{A}\kt$ are normalized eigenvectors of $W=X X^\dagger$.
This implies that the original composite state $|\psi\kt$ attains the form
\begin{equation}
\label{SD}
|\psi\kt = \sum_{i=1}^N \sqrt{\ld_i} |\ld_i^{A}\kt \otimes  | \ld_i^{B}\kt \ 
\end{equation} 
in this diagonal basis. Note that the normalization condition 
$\br \psi | \psi\kt =1$  implies that $\sum_{i=1}^N \ld_i =1$.

Each state $|\ld_i^{A}\kt \otimes |\ld_i^{B}\kt$ is separable in the SD above. However, their linear combination $|\psi\kt$ cannot, in general, be written as a direct product $|\psi\kt = |\phi^{A}\kt \otimes |\phi^{B}\kt$ of two states of the respective subsystems. The state $|\psi\kt$ is therefore in general entangled, and the Schmidt eigenvalues $\ld_i$ can be used to quantify the degree of entanglement (see below for details). 

A sensible way to introduce randomness in this system is to sample those entangled pure states with equal probability, (i.e., 
 according to the uniform measure) over the full Hilbert space. Physically, this corresponds to assuming the minimal amount of \emph{a priori} information about the quantum state under consideration. Mathematically, this implies that
 the coefficients $\{x_{i,j}\}$ in \eqref{decomp} are uniformly distributed on the manifold $\sum_{i,j}|x_{i,j}|^2=1$ - this condition is necessary to enforce normalization of $|\psi\kt$. Therefore, the probability density function of the $N\times M$ matrix $X$ with entries $x_{i,j}$ can be written as
\beq
P(X)\propto \delta\left(\mathrm{Tr}(X X^\dagger)-1\right) \ ,
\eeq
which implies that $X$ is distributed according to a \emph{fixed-trace} ensemble (see \cite{majreview} for an excellent review). Performing a singular value decomposition of $X$ and integrating out the eigenvectors, the jpdf of Schmidt eigenvalues $\ld_i$ turns out to be given by~\cite{Lubkin}
\beq
P^{(FT)}(\bm\lambda)=Z^{-1}_{N,M}    \,
\delta\left(\sum_{i=1}^N \ld_i -1 \right)  \, |\triangle({\bm \ld})|^{\beta}\, \prod_{i=1}^{N}
\ld_i^\alpha\! \,,
\label{jpdf1}
\eeq 
where 
the normalization constant $Z^{-1}_{N,M}$ is given explicitly by
(see, e.g., Ref. \cite{ace,livan})
\beq
\label{norm}
Z^{-1}_{N,M}=\frac{\Gamma(\mu) \Gamma^N(1+\beta/2)}{\prod_{j=0}^{N-1}\Gamma((M-j)\beta/2)\Gamma(1+(N-j)\beta/2)} \,,
\eeq
with $\mu = \beta N M/2$,
$\alpha={\frac{\beta}{2}(b +1)-1}$, $b = M - N \geq 0$. In \eqref{jpdf1}, $\beta = 2$ is the Dyson index and
 $\triangle({\bm \ld}) = \prod_{j<k} (\ld_j-\ld_k)$ is the Vandermonde determinant. In what follows averaging with respect to the distribution in \eqref{jpdf1} will be denoted by $\mathbb{E}_{FT}\left\{ \ldots \right\}$.

\section{$n$-point densities and cross-moments of the FT ensemble}
\label{densities}

Going beyond the linear statistics requires the knowledge 
of the $n$-point densities of the fixed-trace ensemble in \eqref{jpdf1},
\beq
\label{r}
{\rho}^{(FT)}_n([\bm\lambda]_n)=\int_0^\infty  P^{(FT)}(\bm\lambda)\prod_{k= n+1}^N d\lambda_{k} \ .
\eeq 

In principle, this analysis has been performed in the recent paper \cite{liu}, which focused however on a particular scaling limit so that extracting ${\rho}_n^{(FT)}([\bm\lambda]_n)$
from their general formulae is not that easy. 
On the other hand, we realize that the derivation of  ${\rho}_n^{(FT)}([\bm\lambda]_n)$ with arbitrary $n$, $N$ and $M$ is rather straightforward and moreover, sheds some light on the physical meaning of the Schmidt eigenvalues. Thus for the sake of completeness we present it here. We also 
note that although $\beta = 2$ in quantum context, such a derivation can be done for any  value of $\beta$ in \eqref{jpdf1}.

\subsection{Relation between the $n$-point densities of the $\beta$-FT and the $\beta$-WL ensembles }

Let $P^{(WL)}({\bm y})$ 
and ${\rho}^{(WL)}_n([{\bm y}]_n)$
denote the jpdf and 
the normalized $n$-point densities (with $n \leq N$) of the eigenvalues $y_i$, $i = 1,2,\ldots,N$, 
of the $\beta$-WL ensemble, respectively 
\beq
P^{(WL)}({\bm y})\!= \frac{Z^{-1}_{N,M}}{2^{\mu} \Gamma(\mu)}  \,  |\triangle({\bm y})|^{\beta}  \prod_{i=1}^{N}
y_i^\alpha\! \, 
e^{ - y_i/2}\,,
\label{jpdf2}
\eeq 
and 
\beq
{\rho}^{(WL)}_n([{\bm y}]_n)=\int_0^\infty P^{(WL)}({\bm y}) \prod_{k = n+1}^N dy_{k} \,.
\eeq
Averaging with respect to the distribution in \eqref{jpdf2} will be denoted in what follows by the symbol $\mathbb{E}_{WL}\left\{ \ldots \right\}$.

Introducing 
the Laplace transform of a function $f(t)$ 
\begin{equation}
\tilde{f}(p) = {\cal L}_{p,t}\left(f(t)\right) = \int_0^\infty dt \, e^{- p t} f(t) \,,
f(t) = {\cal L}^{-1}_{t,p}\left(\tilde{f}(p)\right) \,,
\end{equation}
the general relation between the $n$-level densities of the $\beta$-fixed-trace and the $\beta$WL ensembles can be 
obtained using the following standard approach. We define first 
two auxiliary functions
\beq
P^{(FT)}({\bm\lambda};t)\!=Z^{-1}_{N,M}    \,
\delta\left(\sum_{i=1}^N \ld_i -t \right)  \, |\triangle({\bm \ld})|^{\beta}\, \prod_{i=1}^{N} \,
\ld_i^\alpha\! \,,
\label{jpdf3}
\eeq 
and
\beq
\label{rt}
{\rho}^{(FT)}_n([{\bm\lambda}]_n;t)= \int_0^\infty  P^{(FT)}({\bm\lambda};t) \prod_{k = n+1}^N d\lambda_{k} \,,
\eeq
which are mere generalizations of the expressions in  \eqref{jpdf1} and  \eqref{r}
for the case of a trace fixed to be equal to $t > 0$. Note that $P^{(FT)}({\bm\lambda};t)$ is a normalized joint pdf only for $t =1$.

Taking now the Laplace transform of ${\rho}^{(FT)}_n([{\bm\lambda}]_n;t)$, we have 
\begin{align}
\label{rt}
& \tilde{\rho}^{(FT)}_n([{\bm\lambda}]_n;p)= 
\int_0^\infty  \tilde{P}^{(FT)}({\bm\lambda};p) \prod_{k = n+1}^N d\lambda_{k} \nonumber\\
&= Z^{-1}_{N,M}  \, \int_0^\infty   |\triangle({\bm \ld})|^{\beta} \prod_{k = n+1}^N d\lambda_{k} \,  \prod_{i=1}^{N} \,
\ld_i^\alpha\! \, e^{- p \,  \lambda_i } \,.
\end{align}
Further on, changing the integration variables $\lambda_i \to y_i/2 p$, we formally rewrite the latter equation as 
\begin{align}
\label{rt1}
\tilde{\rho}^{(FT)}_n\left(\frac{[\bm y]_n}{2 p};p\right)&=
 \frac{Z^{-1}_{N,M}}{( 2 p )^{\mu - n}} 
\int_0^\infty \prod_{k = n+1}^N d\lambda_{k}  \, |\triangle({\bm y})|^{\beta}\, \nonumber\\
&\times \prod_{i=1}^{N} \,
y^\alpha\!  \, e^{-  y_i/2}  \,,
\end{align}
from which we read off the following relation 
\begin{equation}
\label{general}
\tilde{\rho}^{(FT)}_n([\bm\lambda]_n;p) \equiv \frac{2^n \, \Gamma(\mu)}{p^{\mu - n}}  \, {\rho}^{(WL)}_n(2 p [\bm\lambda]_n) \ .
\end{equation}
This relations holds for arbitrary $\beta$, $n$, $N$ and $M$. Inversion of the Laplace transform yields 
\begin{equation}
\label{central}
{\rho}^{(FT)}_n([{\bm\lambda}]_n;t) = 2^n \, \Gamma(\mu) \, {\cal L}^{-1}_{t,p}\left(\frac{{\rho}^{(WL)}_n(2p[\bm\lambda]_n)}{p^{\mu - n}}\right) \,,
\end{equation}
and the desired general relation between the $n$-point densities of the two ensembles follows setting $t=1$
\begin{equation}
\label{central9}
{\rho}^{(FT)}_n([\bm\lambda]_n) = 2^n \, \Gamma(\mu) \, {\cal L}^{-1}_{t=1,p}\left(\frac{{\rho}^{(WL)}_n(2p[\bm\lambda]_n)}{p^{\mu - n}}\right) \,.
\end{equation}
Hence, since ${\rho}^{(WL)}_n([{\bm y}]_n)$ are known from the general theory of orthogonal polynomials in terms of $n\times n$ determinants of a kernel built out of Laguerre polynomials, 
the $n$-level densities for fixed-trace ensembles for arbitrary $n$ are obtained by the  
inversion of the Laplace transform with respect to $p$, upon setting $t = 1$.

\subsection{$\beta$-FT vs scaled-variables ensemble}

It may be instructive
to re-derive the result in \eqref{central9} using a different approach, which sheds some light on the physical meaning of the Schmidt eigenvalues.  We focus on the $\beta$-WL ensemble \eqref{jpdf2} and introduce \emph{scaled variables} (see, e.g., \cite{krish,wei2,nadler}) of the form
\begin{equation}
\label{defomega}
\overline{\omega}_i = N \omega_i = \frac{y_i}{N^{-1} \sum_{i = 1}^N y_i} \,, \omega_i \in [0,1]\,.
\end{equation}
Note that such variables automatically obey the fixed-trace constraint, $\sum_{i = 1}^N \omega_i \equiv 1$. The physical significance
of $\overline{\omega}_i$ is evident:  it measures an individual $y_i$ 
against the arithmetic mean eigenvalue 
in a given 
realization of the $\beta$-WL ensemble. Consequently, the one-point density of $\overline{\omega}_i$ 
shows how heterogeneous the distribution of the eigenvalues $y_i$ is and how likely it is that $y_i$-s concentrate around their mean value.

Note that the extreme value statistics of such random variables (the largest, $\omega_{\rm max} =  y_{\rm max} /\sum_{i = 1}^N y_i$, and the smallest, $\omega_{\rm min} =  y_{\rm min}/\sum_{i = 1}^N y_i$, which is the reciprocal of the so called Demmel condition number, see, e.g., \cite{edel} and more recent \cite{wei1,mac}) plays a key role in various  
scale independent hypothesis testing procedures, both in classical statistics as well as in signal processing. Classical examples (see, e.g., \cite{wei2,nadler}) include testing for the presence of interactions in multi-way data and testing for equality of the population covariance to a scaled identity matrix. Modern signal processing applications include testing for the presence of signals in cognitive radio as well as non-parametric signal detection in array processing. Spectral densities of ordered 
$\omega_{i}$ have been 
determined long time ago \cite{krish}, while analogous distributions for unordered, random 
scaled variables have been
recently evaluated for the $\beta$-WL ensemble for small $N$ \cite{greg} and the Gaussian Unitary Ensemble for arbitrary $N$ \cite{mauricio}. We also note that such scaled variables have been also used  to characterize 
the effective broadness of "narrow" distributions possessing moments of arbitrary order \cite{carlos,thiago}.      

Using \eqref{jpdf2} and taking into account the well-known fact that the distribution $F(t)$ of
the trace $t = \sum_{i=1}^N y_i$ 
in the $\beta$-WL ensemble is the central $\chi^2$-distribution of the form
\begin{equation}
\label{F}
F(t) = \frac{t^{\mu - 1}}{2^{\mu} \Gamma\left(\mu\right)} \, e^{- t/2} \,,
\end{equation}
one finds
the jpdf $\Psi(\bm\omega)$
of
the  
variables $\omega_i$ (see, e.g., \cite{krish})
\begin{equation}
\label{psi}
\Psi(\bm\omega)=Z^{-1}_{N,M}    \,
\delta\left(\sum_{i=1}^N \omega_i -1 \right)  \, |\triangle({\bm \omega})|^{\beta}\, \prod_{i=1}^{N}
\omega_i^\alpha\! \,.
\end{equation}
Remarkably, (but not counter-intuitively), the scaled variables $\omega_i$ appear to have exactly the same jpdf as the Schmidt eigenvalues.
This means, in particular, that the distributions of the largest  and the smallest  scaled variables $\omega_{\rm max}$ and $\omega_{\min}$ coincide with the distribution of the largest and the smallest fixed-trace eigenvalue, and also implies that the $n$-point densities $\psi_n([\bm\omega]_n)$ have the same functional form as the $n$-level densities ${\rho}^{(FT)}_n([\bm\lambda]_n)$.

Next, we formally represent ${\rho}^{(WL)}_n([{\bm y}]_n)$ as
\begin{align}
{\rho}^{(WL)}_n([{\bm y}]_n)&=\int F(t) dt \int  \psi_n([\bm\omega]_n) \prod_{k = n+1}^N d\omega_{k} \, \nonumber\\
&\times \delta\left(y_k - t \omega_k \right) \,,
\end{align}
which gives, upon integration over the $\omega$-variables, 
\begin{equation}
{\rho}^{(WL)}_n([{\bm y}]_n)=\int \frac{F(t)}{t^n} \, \psi_n\left(\frac{[\bm y]_n}{t}\right) dt\,.
\end{equation}
Using the definition of $F(t)$ in \eqref{F}, changing the integration variable $t \to 2 p t$ and also setting 
$y_i = 2 p \omega_i$, we can cast the latter equation into the form
\begin{equation}
{\rho}^{(WL)}_n(2p[\bm\omega]_n)= \frac{p^{\mu - n}}{2^n \Gamma(\mu)} \, {\cal L}_{p,t}\left( \psi_n\left(\frac{[\bm\omega]_n}{t}\right)\right)  \,,
\end{equation}
which yields straightforwardly 
\begin{equation}
\label{central2}
\psi_n\left(\frac{[\bm\omega]_n}{t}\right) = 2^n \Gamma(\mu) \, {\cal L}^{-1}_{t,p}\left( \frac{{\rho}^{(WL)}_n(2p[\bm\omega]_n)}{p^{\mu - n}}\right)  \,.
\end{equation}
Our previous result in \eqref{central9} follows directly from \eqref{central2} by setting $t=1$. 

\subsection{Cross-moments of the $\beta$-FT and $\beta$-WL ensembles}

Further on, our aim is now to establish a relation between the cross-moments of the $\beta$-FT and $\beta$-WL ensembles.
To this end, we first note that, evidently,
\begin{equation}
\label{c1}
\mathbb{E}_{FT}\left\{ \lambda_1^{a_1} \lambda_2^{a_2} \ldots \lambda_n^{a_n} \right\} = \mathbb{E}_{\bm\omega}\left\{ \omega_1^{a_1} \omega_2^{a_2} \ldots \omega_n^{a_n}\right\} \,,
\end{equation}
where the symbol $\mathbb{E}_{\bm\omega}\{\ldots\}$ on the right-hand-side 
denotes averaging with respect to the jpdf of the scaled variables in \eqref{psi}.
Further on, using the definition of the scaled variables in \eqref{defomega}, we write 
\begin{align}
\label{c2}
&\mathbb{E}_{\bm\omega}\left\{ \omega_1^{a_1} \omega_2^{a_2} \ldots \omega_n^{a_n}\right\} = \frac{Z_{N,M}^{-1}}{2^{\mu} \Gamma(\mu)} \nonumber\\
&\times\int \frac{\prod_{i=1}^n y_i^{a_i}}{\left(\sum_{i=1}^N y_i\right)^{\sum_{i=1}^n a_i}}  |\triangle({\bm \ld})|^{\beta} \, \prod_{i=1}^N y_i^{\alpha} e^{-y_i/2} dy_i \ .
\end{align}
Next, taking advantage of the integral identity
\begin{equation}
\int^{\infty}_0 z^{k-1} e^{- p z} dz = \frac{\Gamma(k)}{p^k} \,,
\end{equation}
equation \eqref{c2} can be formally rewritten as
\begin{align}
\label{c3}
&\mathbb{E}_{\bm\omega}\left\{ \omega_1^{a_1} \omega_2^{a_2} \ldots \omega_n^{a_n}\right\} = \frac{Z_{N,M}^{-1}}{2^{\mu} \Gamma(\mu) \Gamma\left(\sum_{i=1}^n a_i\right)} \nonumber\\
&\times \int^{\infty}_0  z^{\sum_{i=1}^n a_i - 1} dz 
  \int \prod_{i=1}^n y_i^{a_i} \, |\triangle({\bm \ld})|^{\beta} \,\nonumber\\
 &\times  \prod_{i=1}^N y_i^{\alpha} e^{-(z+1/2) y_i} dy_i \ .
\end{align}
Changing the integration variables $y'_i \to (1 + 2 z) y_i$, we have, dropping the prime,
\begin{align}
\label{c4}
&\mathbb{E}_{\bm\omega}\left\{ \omega_1^{a_1} \omega_2^{a_2} \ldots \omega_n^{a_n}\right\} = \frac{Z_{N,M}^{-1}}{2^{\mu} \Gamma(\mu) \Gamma\left(\sum_{i=1}^n a_i\right)} \nonumber\\
&\times  \int^{\infty}_0  \frac{z^{\sum_{i=1}^n a_i - 1}}{(1 + 2 z)^{\mu + \sum_{i = 1}^n a_i}} dz  \int \prod_{i=1}^n y_i^{a_i} \, |\triangle({\bm \ld})|^{\beta} \, \nonumber\\
&\times \prod_{i=1}^N y_i^{\alpha} e^{- y_i/2} dy_i \,,
\end{align}
from which we immediately obtain the following relation between the cross-moments of $\lambda_i$-s and $y_i$-s:
\begin{equation}
\label{central3}
\dfrac{\mathbb{E}_{FT}\left\{ \lambda_1^{a_1} \lambda_2^{a_2} \ldots \lambda_n^{a_n}\right\}}{\Gamma(\mu)} = \dfrac{\mathbb{E}_{WL}\left\{ y_1^{a_1} y_2^{a_2} \ldots y_n^{a_n}\right\}}{2^{\sigma_n} \Gamma\left(\mu + \sigma_n\right)}  \,,
\end{equation}
where $\sigma_n = \sum_{i=1}^n a_i$ and 
$\{a_i\}$ are arbitrary (not necessarily integer) numbers each greater than $-b$, 
conditioned by the constraint that their sum $\sigma_n = \sum_{i=1}^n a_i$ is strictly larger than zero. This result is a direct consequence of \eqref{central2} and is valid for arbitrary $\beta$ and arbitrary $n \leq N \leq M$. While preparing this manuscript, we became aware that the same relation between the cross-moments has been recently presented in \cite{atkin}.

Eq. \eqref{central3} entails a series of very useful identities
between the powers of traces of the $\beta$-fixed-trace and the $\beta$-WL ensembles, respectively. 
Multiplying both sides of \eqref{central3} by $\prod_{i=1}^n (-1)^{a_i} p_i^{a_i}/a_i!$ (with all $p_i \geq 0$) and performing summation over all positive integer $a_i$, we get the following relation between the weighted traces of two ensembles 
\begin{align}
\label{hh1}
&\dfrac{\mathbb{E}_{FT}\left\{ \exp\left( - \sum_{i=1}^{n} p_i \ld_i\right) \right\}}{\Gamma(\mu)} = \nonumber\\
&=\mathbb{E}_{WL}\left\{ \dfrac{J_{\mu-1}\left(\sqrt{2 \sum_{i=1}^{n} p_i y_i}\right)}{\left( \sum_{i=1}^{n} p_i y_i\right)^{\frac{(\mu-1)}{2}}} \right\} \ ,
\end{align}
where $J_{\mu-1}(\ldots)$ is a Bessel function, or, equivalently,
\begin{equation}
\mathbb{E}_{FT}\left\{ 
\dfrac{1}{\left(1 +2 \sum_{i=1}^{n} p_i \ld_i\right)^{\mu}} \right\} = \mathbb{E}_{WL}\left\{ \exp\left( - \sum_{i=1}^{n} p_i y_i\right)\right\}\ .
\end{equation} 
Further on, setting $a_i = 2 n_i$, where $n_i$ are positive integers, multiplying both sides of \eqref{central3} by $\prod_{i=1}^n (-1)^{n_i} p_i^{n_i}/n_i!$ and performing summations over all $n_i$, we find the following relation between the weighted squared traces
\begin{align}
\label{aa}
&\mathbb{E}_{FT}\left\{ \exp\left( - \sum_{i=1}^{n} p_i \ld_i^2\right)\right\} = \nonumber\\
&=\mathbb{E}_{WL}\left\{ 
_0F_2\left(\frac{\mu}{2},\frac{\mu+1}{2}, -\frac{1}{16}\sum_{i=1}^{n} p_i y_i^2\right)\right\} \,, 
\end{align}
where $_0F_2(\ldots)$ is the generalized hypergeometric series, and
\begin{align}
\label{bb}
& \mathbb{E}_{FT}\left\{ \dfrac{U\left(\frac{\mu}{2},\frac{1}{2}, \dfrac{1}{4 \sum_{i=1}^{n} p_i \ld_i^2}\right)}{\left(4 \sum_{i=1}^{n} p_i \ld_i^2 \right)^{\mu/2}} \right\} = \nonumber\\
&\mathbb{E}_{WL}\left\{ \exp\left( - \frac{1}{4} \sum_{i=1}^{n} p_i y_i^2\right) \right\} \,,
\end{align}
where $U(\ldots)$ is the Tricomi's confluent hypergeometric function. We note that both equations can be straightforwardly used for the derivation of the moments of purity \eqref{aa} and of its reciprocal value - the Schmidt number \eqref{bb}, which we will demonstrate in Sec. \ref{Schmidt}. Eventually, by taking advantage of our \eqref{aa}, (or by simply using the multinomial theorem and our \eqref{central3}) it is straightforward to show that the moments of the squared weighted traces of two ensembles obey
\begin{equation}
\label{kkk}
\dfrac{\mathbb{E}_{FT}\left\{\left(\sum_{i=1}^n p_i \ld_i^2\right)^{a} \right\}}{\Gamma(\mu)} = \dfrac{\mathbb{E}_{WL}\left\{\left(\sum_{i=1}^n p_i y_i^2\right)^{a} \right\}}{2^{2 a}\Gamma(\mu + 2 a)} \,,
\end{equation} 
where $a$ is an arbitrary positive integer and $p_i$ are arbitrary non-negative numbers. 
More generally, for the exponential moments of the 
$q$-parametrized 
R\'enyi entropy, we have
\begin{equation}
\label{hh2}
\dfrac{\mathbb{E}_{FT}\left\{\left(\sum_{i=1}^n p_i \ld_i^q\right)^{a} \right\}}{\Gamma(\mu)} = \dfrac{\mathbb{E}_{WL}\left\{\left(\sum_{i=1}^n p_i y_i^q\right)^{a} \right\}}{2^{q a}\Gamma(\mu + q a)} \,,
\end{equation} 
where $q$ is an arbitrary (not necessarily integer) positive number. We emphasize that our equations \eqref{hh1} to \eqref{hh2} are valid for arbitrary $\beta$ and arbitrary $n \leq N \leq M$.

\section{One and two-point densities for the FT ensemble with $\beta=2$}
\label{spec}

In this section we present explicit results for the one and two-point densities of the FT Wishart ensemble at $\beta=2$. 
We remark that the one-point density has been already 
determined for finite $N,M$ in earlier Refs.\cite{theo2,theo3,theo5,theo4} also for $\beta=1,4$. Here we present it (for $\beta=2$) for the sake of completeness, and also to demonstrate that our general relation in \eqref{central9} yields an explicit expression for the spectral density in a most immediate manner, as compared to previous approaches, especially the one in \cite{theo3}, which relied on a quite complicated analytical approach based on the theory of the holonomic system of differential equations. 

\subsection{Spectral density for the FT ensemble}

\begin{figure}[hb]
\centerline{\includegraphics[width=8cm]{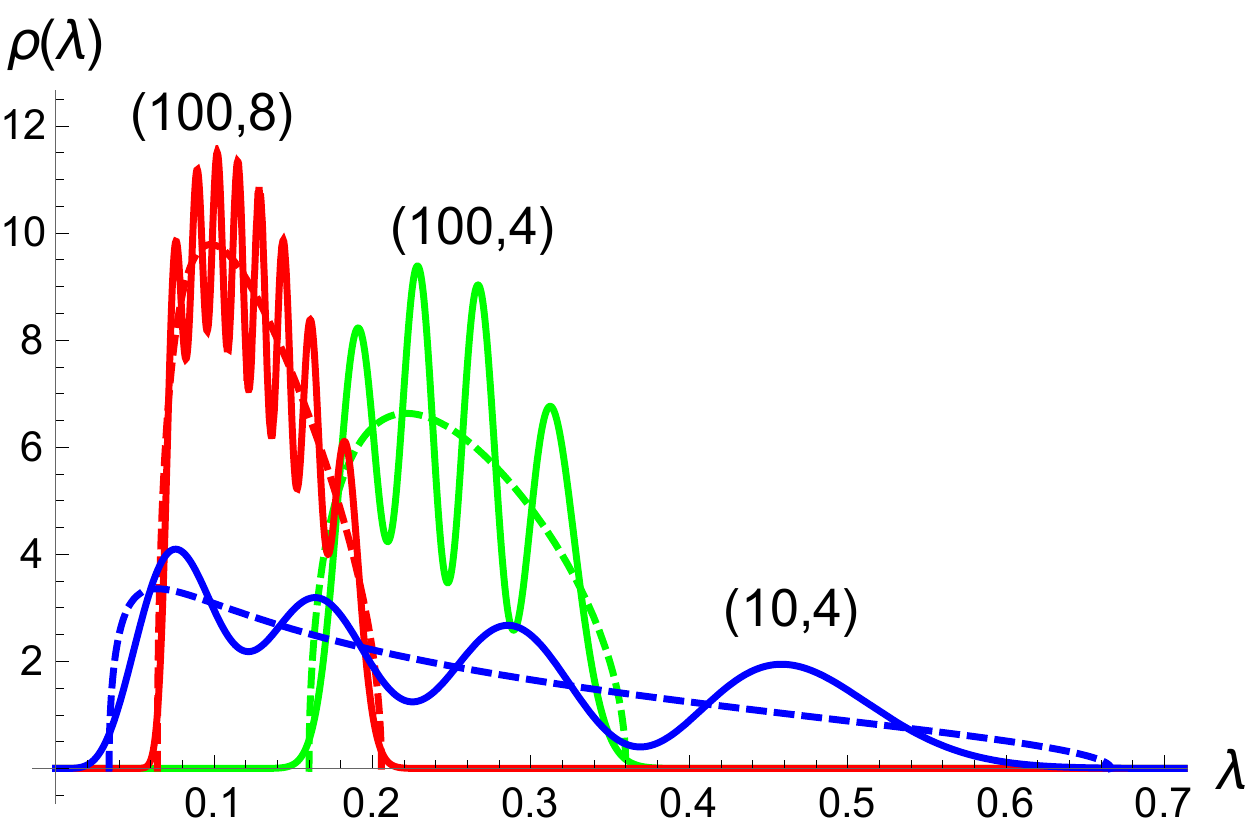}}
\caption{(color online) The spectral density ${\rho}^{(FT)}_1(\ld)$ for the unitary fixed-trace ensemble in \eqref{lambda} for $M =100$ and $N = 8$ (red), 
$M = 100$ and $N = 4$ (green) and $M=10$ and $N = 4$ (blue). The dashed lines define the 
Mar\v{c}enko-Pastur-type distribution in \eqref{pm} corresponding to these values of $M$ and $N$.}
\label{fig1}
\end{figure}

Before we proceed with exact calculations, the following simple observation is in order: 
note that in the limit $\mu \to \infty$, our \eqref{central3} implies that to leading order in $\mu$ and for arbitrary positive $a_1$, one has
\begin{equation}
\mathbb{E}_{FT}\left\{ \lambda^{a_1} \right\} \simeq \mathbb{E}_{WL}\left\{ \left(\frac{y}{2 \mu} \right)^{a_1}\right\} \,.
\end{equation}
This means, in turn, that the Schmidt eigenvalues, as well as the scaled variables $\omega$, have the same limiting distribution as the eigenvalues $y$ of the $\beta=2$ WL ensemble divided by $2 \mu$; that being, they obey the Mar\v{c}enko-Pastur distribution \cite{mar} for $N,M\to\infty$ (with $N/M$ fixed) of the form:
\begin{equation}
\label{pm}
\rho^{(FT)}_1(\lambda) \to \frac{M}{2 \pi} \frac{\sqrt{(B-\lambda) (\lambda - A)}}{\lambda} \,,
\end{equation}
where the boundaries of the support are given by
\begin{equation}
A = \frac{1}{N} \left(1 - \sqrt{\frac{N}{M}}\right)^2 \,,
\end{equation}
and
\begin{equation}
B = \frac{1}{N} \left(1 + \sqrt{\frac{N}{M}}\right)^2 \,.
\end{equation}
The asymptotic result in \eqref{pm} is depicted in Fig.\ref{fig1} together with the exact results for the spectral density.
Note that the result in \eqref{pm} has been obtained earlier in several papers (see, e.g.,  \cite{Znidaric}).

Now, the one-point density for the unitary ($\beta=2$) WL ensemble at fixed $N,M$ can be written as
\begin{equation}
\label{j}
{\rho}^{(WL)}_1(y) = y^{b} e^{- y/2} \sum_{l=0}^{2 N - 2} A^{(1)}_l(N,M) \, y^l \,,
\end{equation}  
where $A_l^{(1)}(N,M)$ are numerical coefficients, which we define in explicit form in Appendix \ref{appendix} (see \eqref{a1}).
Inserting \eqref{j} into \eqref{central9}, we get straightforwardly
\begin{align}
\label{lambda}
&{\rho}^{(FT)}_1(\ld) = 2^{b + 1} \Gamma(\mu) \lambda^{b} \sum_{l=0}^{2 N - 2} A_l^{(1)}(N,M) \, (2 \lambda)^l \nonumber\\
&\times {\cal L}^{-1}_{t=1,p}\left(p^{-\mu + b+ 1 + l} \, e^{ - p \lambda}\right) \nonumber\\
&= 2 \Gamma(\mu) (1 - \lambda)^{\mu - 2} 
 \sum_{l=0}^{2 N - 2} \frac{A_l^{(1)}(N,M)}{\Gamma(\mu - b - l - 1)} \left(\frac{2 \lambda}{1 - \lambda}\right)^{l+b} \,,
\end{align}
where for the $\beta=2$ ensemble $\mu = N M$.
Note that this is precisely the result obtained previously in \cite{theo2,theo3,theo4} by using different approaches. In Fig.\ref{fig1} we plot ${\rho}^{(FT)}_1(\ld)$ in \eqref{lambda} for several values of $M$ and $N$.

\subsection{Two-point densities for the FT ensemble}

\begin{figure*}[t]
\centering
\includegraphics[width=3.5in]{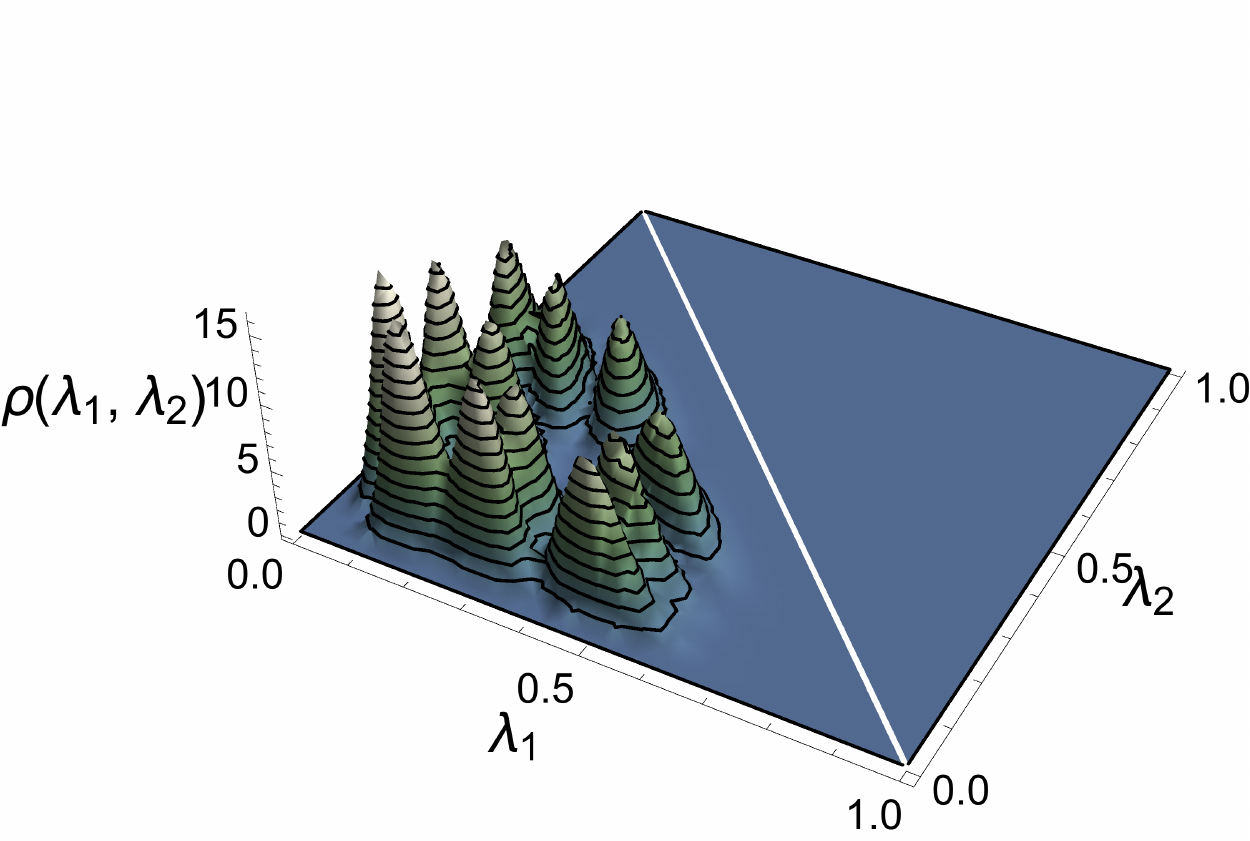}
\includegraphics[width=2.2in]{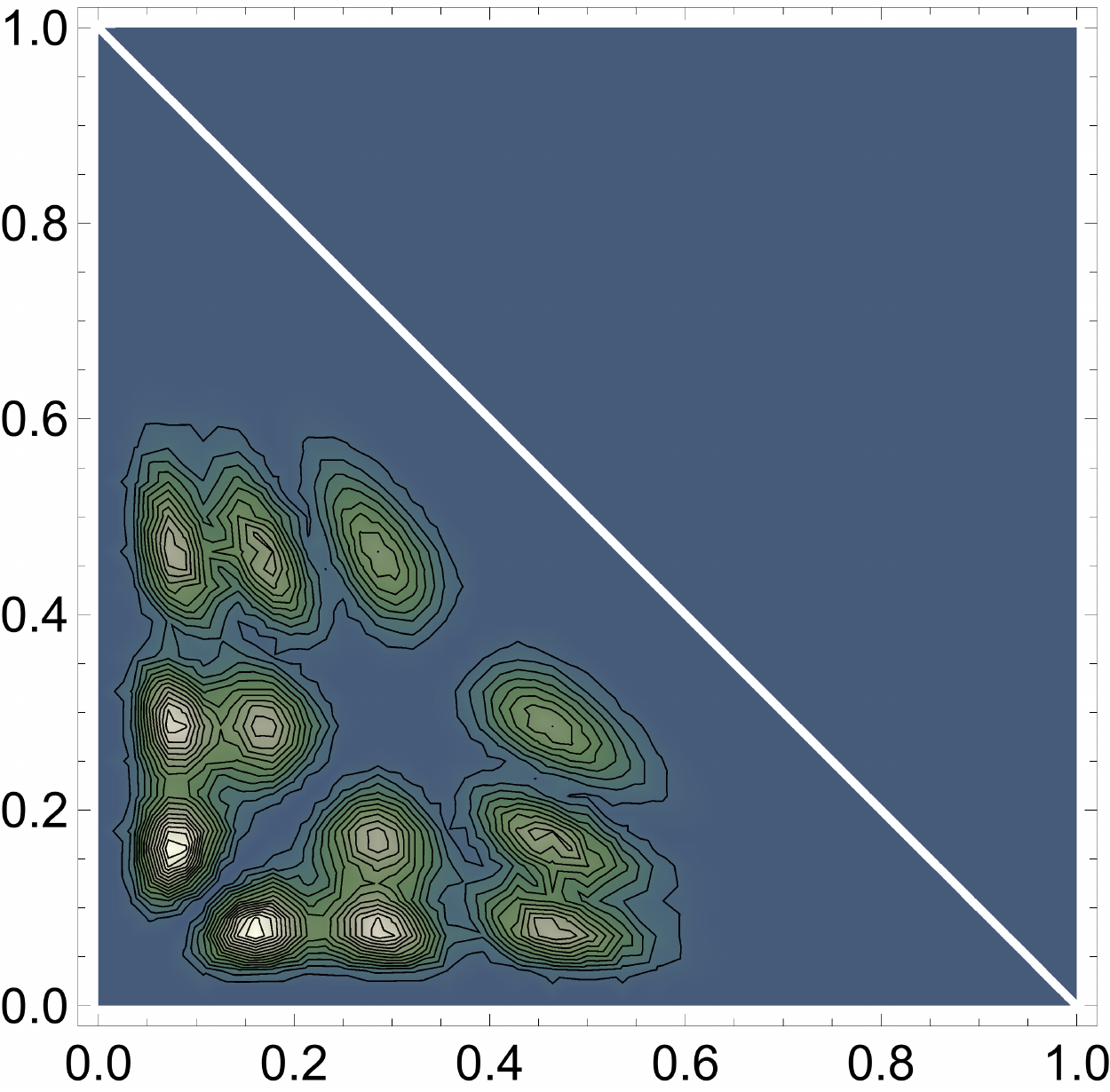}
\includegraphics[width=3.5in]{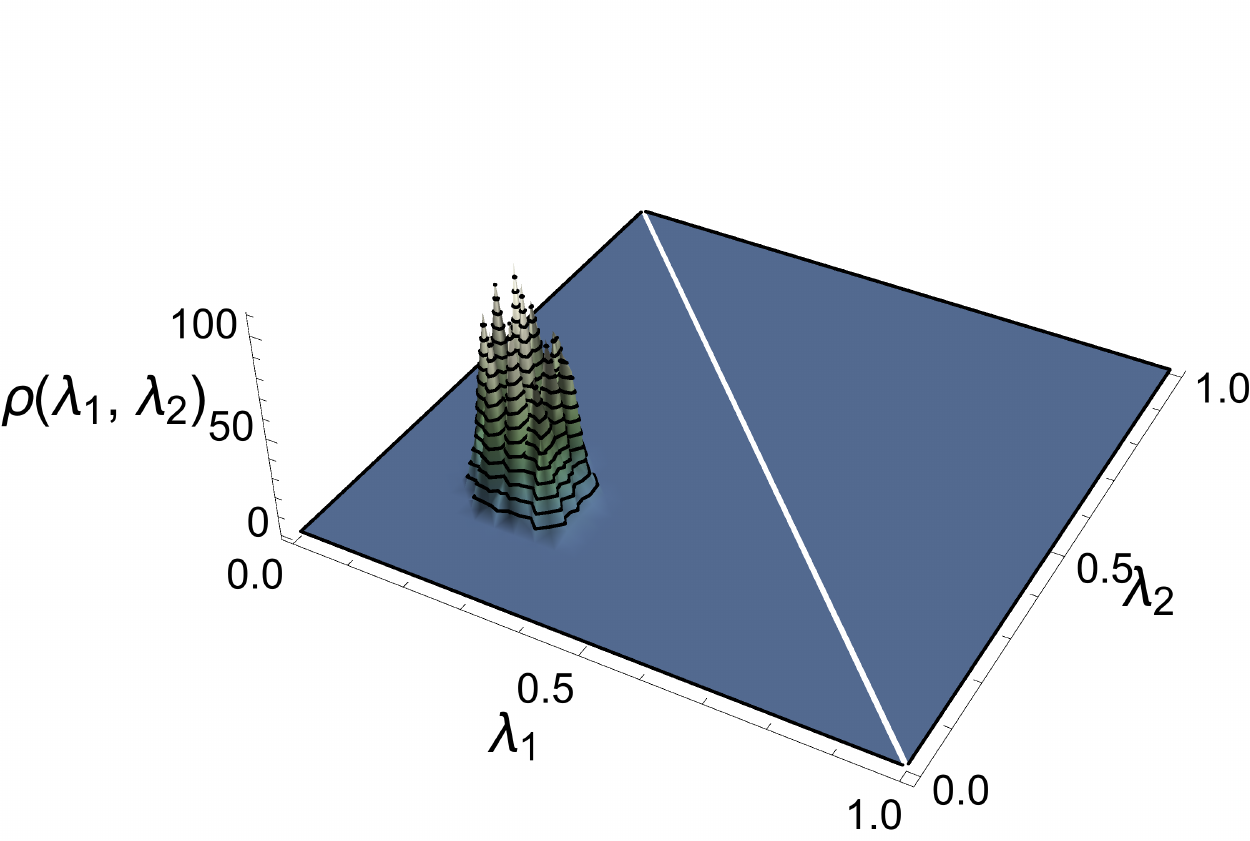}
\includegraphics[width=2.2in]{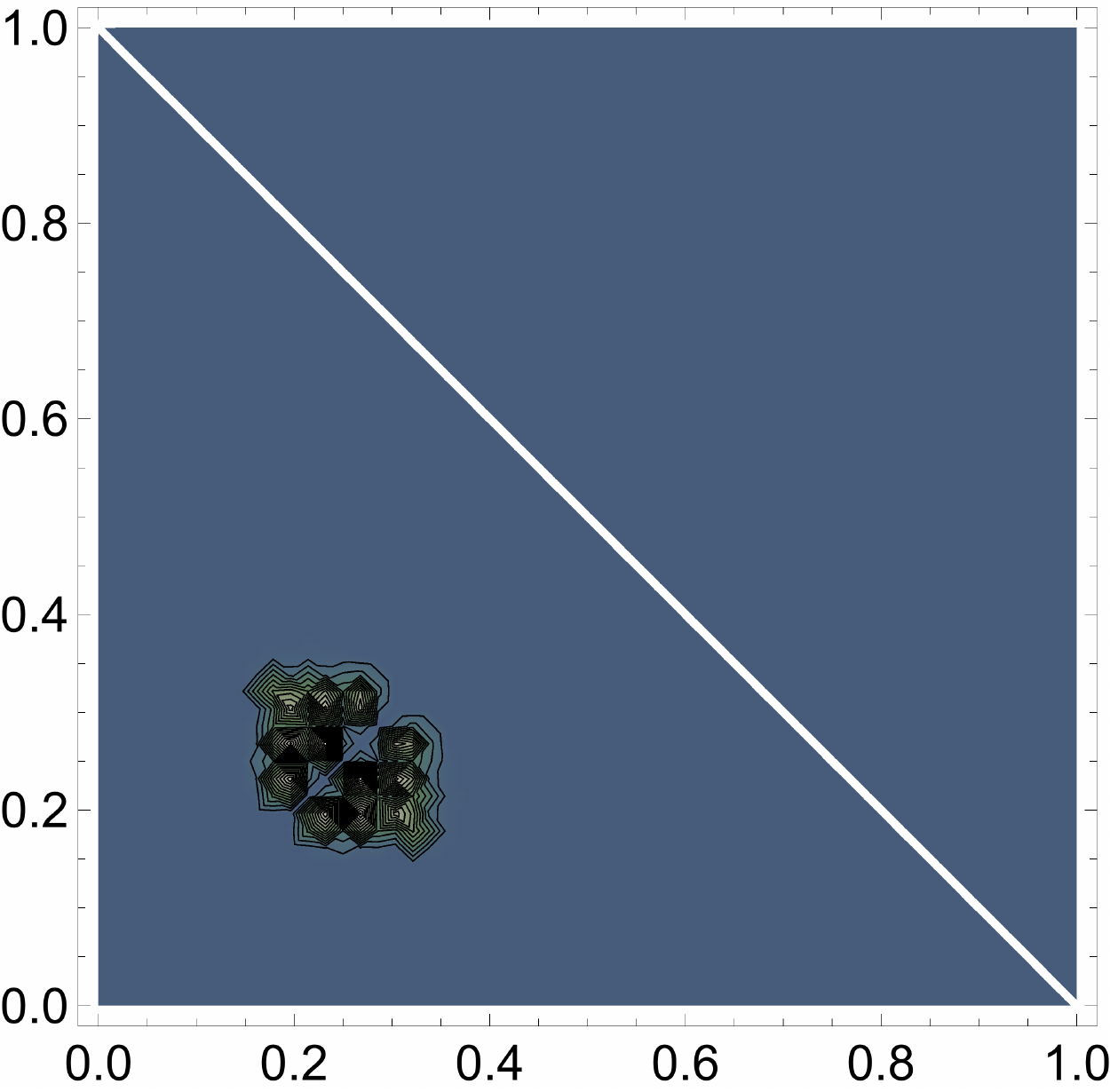}
\caption{(color online) The two-level density ${\rho}^{(FT)}_2(\ld_1,\ld_2)$ for the fixed-trace ensemble
 in \eqref{zz} versus $\ld_1$ and $\ld_2$.  Top: $M =10$ and $N = 4$ - side view (left) and top view (right). 
 Bottom:
$M = 100$ and $N = 4$ - side view (left) and top view (right). }
\label{fig2}
\end{figure*}

The two-level density  ${\rho}^{(WL)}_2(y_1,y_2)$ for the $\beta=2$ Wishart-Laguerre ensemble
 has the following form :
\begin{align}
\label{jj}
{\rho}^{(WL)}_2(y_1,y_2) &= \left(y_1 y_2\right) ^{b} e^{- (y_1 + y_2)/2} \nonumber\\
&\times \sum_{l,j=0}^{2 N - 2} A^{(2)}_{l,j}(N,M) \, y_1^l \, y_2^j \,,
\end{align}  
where $A_{l,j}^{(2)}(N,M)$ are numerical coefficients which are  defined explicitly in Appendix \ref{appendix} (see \eqref{a2}).
Consequently, using \eqref{central9}, the two-level density ${\rho}^{(FT)}_2(\ld_1,\ld_2)$ of the fixed-trace ensemble can be represented in form of the inverse Laplace transform of the following function :
\begin{align}
\label{zzz}
&{\rho}^{(FT)}_2(\ld_1,\ld_2) = 4^{b + 1} \Gamma(\mu) \left(\lambda_1 \lambda_2\right)^{b} 
  \sum_{l,j=0}^{2 N - 2} A_{l,j}^{(2)}(N,M)  \nonumber\\
&\times (2 \lambda_1)^l (2 \lambda_2)^j {\cal L}^{-1}_{t=1,p}\left(p^{-\mu + 2 b + 2 + l + j} e^{ - p (\lambda_1 + \lambda_2)} \right) \,.
\end{align}
Performing the inverse Laplace transform and setting $t=1$, 
we obtain the following exact result 
\begin{align}
\label{zz}
&{\rho}^{(FT)}_2(\ld_1,\ld_2)=  4 \Gamma(\mu) (1 - \lambda_1 - \lambda_2)^{\mu - 3}  \theta(1 - \lambda_1 - \lambda_2)  \nonumber\\
&\times \sum_{l,j=0}^{2 N - 2} \frac{A_{l,j}^{(2)}(N,M)}{\Gamma(\mu - 2 b - l - j -  2)} \frac{(2 \lambda_1)^{l+b} (2 \lambda_2)^{j+b}}{(1 - \lambda_1 - \lambda_2)^{j + l + 2 b}} \ ,
\end{align}
where $\theta(\cdot)$ is the Heaviside step function.  To the best of our knowledge, this is a new result.
The two-level density in \eqref{zz} is plotted in Fig.\ref{fig2} for two different values of $M$ and fixed $N = 4$ and reveals a pronounced structuring and correlations between the eigenvalues.

Finally, we present an explicit expression for the pair cross-moments of arbitrary order
\begin{align}
\label{moments}
&\mathbb{E}_{FT}\left\{ \ld_1^{a_1} \ld_2^{a_2}\right\} = \frac{4^{1+b} \Gamma(\mu)}{\Gamma(\mu + a_1+a_2)}  
\sum_{l,j=0}^{2 N - 2} 2^{l+j} A_{l,j}^{(2)}(N,M) \nonumber\\
&\times \Gamma(1+l+b+a_1)  \Gamma(1+j+b+a_1) \,,
\end{align}
which holds for arbitrary (not necessarily integer and positive) $a_1 > - b - 1$ and $a_2 > - b - 1$, as well as for arbitrary $N$ and $M$.

\section{The variance of the von Neumann entropy }
\label{var}

The average von Neumann entropy can be straightforwardly calculated 
using our \eqref{lambda} to give
 \begin{align}
 \label{mean}
&\mathbb{E}_{FT}\left\{ S_{\rm vN} \right\} = - N \mathbb{E}_{FT}\left\{ \lambda \ln(\lambda)\right\} = \nonumber\\
& \frac{2^{b+1}}{M} \sum_{l=0}^{2 N - 2} 2^l \Gamma(l + b + 2) A_l^{(1)}(N,M) \sum_{k = l + b +2}^{\mu} \frac{1}{k}\,,
\end{align}
which holds for any $M$and $N$ (with $M \geq N$). It is straightforward to verify, e.g., numerically, that the latter expression coincides
with the one conjectured by Page \cite{Page}
\begin{align}
\label{mean2}
&\mathbb{E}_{FT}\left\{ S_{\rm vN} \right\} = \sum_{k = M}^{\mu} \frac{1}{k} - \frac{N+1}{2 M} \nonumber\\
&= \psi^{(0)}(N M + 1) - \psi^{(0)}(M) - \frac{N+1}{2 M} \,,
\end{align}
where $\psi^{(0)}(z) = d \ln \Gamma(z)/dz$ is the digamma function. This expression 
was subsequently proven in \cite{theo1}. Note that $\mathbb{E}_{FT}\left\{ S_{\rm vN} \right\}$ is a monotonically
 \textit{increasing} function of both $M$, (with a fixed $N$), and $N$, (at a fixed $M \geq N$). In both cases, to leading order in $N$, the average von Neumann entropy
 $\mathbb{E}_{FT}\left\{ S_{\rm vN} \right\} \simeq \ln N$, which is precisely 
 the reason to argue that the system is close to a maximally entangled state. 
 
 We note here that there is, however, some subtlety concerning the degree of the entanglement when this issue is analyzed in terms of the entanglement entropy, which is a logarithmic function of $N$. We proceed to show in Section \ref{Schmidt} that the situation is, as a matter of fact, more delicate. Studying the behavior of the moments of the Schmidt number $ K$,  we realize that $K \sim N$ (so that the complete entanglement is achieved) only for the situations when a subsystem of some fixed size $N$ is coupled to a thermodynamically large bath with $M \to \infty$. On the contrary, when the system is partitioned in two subsystems of equal size $N$, $K$ approaches the value $N/2$ only, when $N \to \infty$. Consequently, such square systems are far from being completely entangled.

Next, the variance of the von Neumann entropy 
is by definition
\begin{align}
\label{2}
&{\rm Var}\left(S_{\rm vN}\right) = \mathbb{E}_{FT}\left\{ S^2_{\rm vN} \right\}  - \mathbb{E}_{FT}^2\left\{ S_{\rm vN} \right\} = \nonumber\\
&N \left(\mathbb{E}_{FT}\left\{ \lambda^2 \ln^2(\lambda) \right\} + 
 (N - 1) \mathbb{E}_{FT}\left\{ \lambda_1 \ln(\lambda_1) \lambda_2 \ln(\lambda_2) \right\} \right) \nonumber\\
&- \left(\sum_{k = M}^{\mu} \frac{1}{k} - \frac{N+1}{2 M}\right)^2\,.
\end{align}
To perform averaging, we may proceed in three equivalent but different ways: we may directly use the one and two-point densities in our  \eqref{lambda} and \eqref{zz}, apply our  \eqref{moments} and use the usual replica trick, or taking advantage 
of our \eqref{moments}, express 
$\langle \lambda^2 \ln^2(\lambda) \rangle_{FT}$ and $\langle \lambda_1 \ln(\lambda_1) \lambda_2 \ln(\lambda_2) \rangle_{FT}$ via analogous moments of the unitary WL ensemble and perform averaging using the spectral and two-level densities of the unitary WL ensemble. Here we follow the first approach.

Using our \eqref{lambda} we readily find that the first term on the right-hand-side of \eqref{2} is given by
\begin{align}
\label{first}
&\mathbb{E}_{FT}\left\{ \lambda^2 \ln^2(\lambda) \right\} = \frac{2^{b+1}}{\mu (\mu+1)} \sum_{l = 0}^{2 N - 2} 2^l \Gamma(l + b + 3) \nonumber\\ &\times A_l^{(1)}(N,M) 
  \left(\left(\sum_{k = l + b + 3}^{\mu + 1} \frac{1}{k}\right)^2 + \sum_{k=l+b+3}^{\mu+1} \frac{1}{k^2}\right) \,.
\end{align}
In a similar fashion, we find that the cross term in \eqref{2} obeys
\begin{align}
\label{second}
&\mathbb{E}_{FT}\left\{ \lambda_1 \ln(\lambda_1) \lambda_2 \ln(\lambda_2) \right\} = \frac{4^{1 + b}}{\mu (\mu + 1)} \sum_{l,j=0}^{2 N - 2} 2^{l+j}   \nonumber\\
&\times \Gamma(l + b + 2) \Gamma(j + b + 2) A_{l,j}^{(2)}(N,M) \nonumber\\
&\times  \left(\sum_{k=l+b+2}^{\mu + 1} \frac{1}{k} \sum_{k'=j+b + 2}^{\mu + 1}\frac{1}{k'} - \sum_{k= \mu + 2}^{\infty}\frac{1}{k^2}\right) \,.
\end{align}
Equations \eqref{2}, \eqref{first} and \eqref{second} define an exact expression for the variance of the von Neumann entropy, which is valid for arbitrary $N$ and $M$, and for any particular choice of $N$ and $M$ it can be readily evaluated using Mathematica. On the other hand, it has quite a complicated structure so that its dependence on $N$ and $M$ can not be easily understood. Based on low $N,M$ evaluations, we realize eventually that the formula for the variance of the entropy can be cast into a much more compact exact form
\begin{align}
\label{N}
{\rm Var}\left(S_{\rm vN}\right) &= - \psi^{(1)}(N M+1) +  \frac{M + N}{N M + 1} \psi^{(1)}(M) \nonumber\\
&- \frac{(N+1) \left(N + 1 + 2 M\right) }{4 M^2 (N M + 1)}\ ,
\end{align}
where $\psi^{(1)}(z) = d^{2} \ln \Gamma(z)/d z^{2}$ is the trigamma function, which is defined  
for integer values of the argument
as a truncated sum of the form  $\psi^{(1)}(n) = \sum_{k=n}^{\infty} 1/k^2$.

\begin{figure}[ht!]
\begin{center}
\includegraphics[width=8cm]{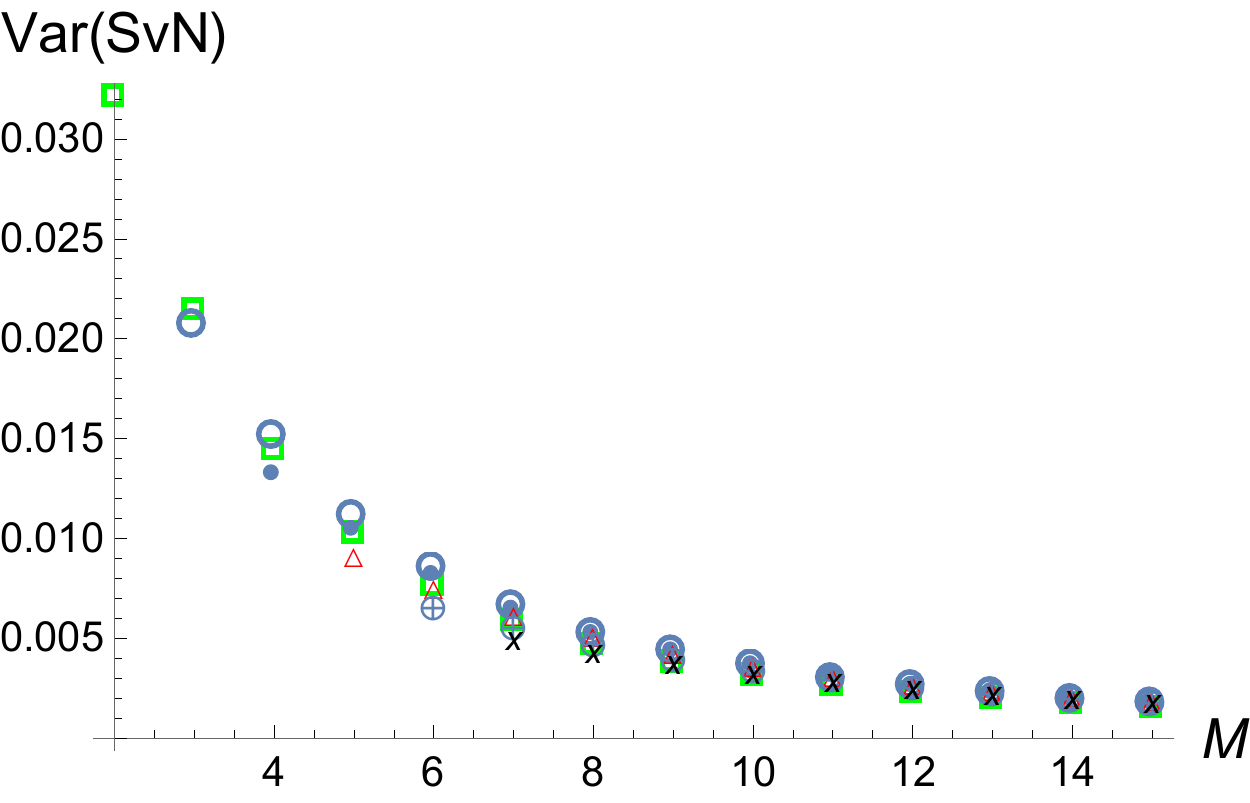}
\end{center}
\caption{(color online) The variance ${\rm Var}\left(S_{\rm vN}\right)$ in \eqref{N}, 
as a function of $M$
for $N = 2$ (squares), $N = 3$  (open circles), $N=4$ (filled circles), $N = 5$ (triangles), $N=6$ (crossed triangles) and $N=7$ (crosses).}
\label{fig4}
\end{figure}

\begin{figure}[ht!]
\begin{center}
\includegraphics[width=8cm]{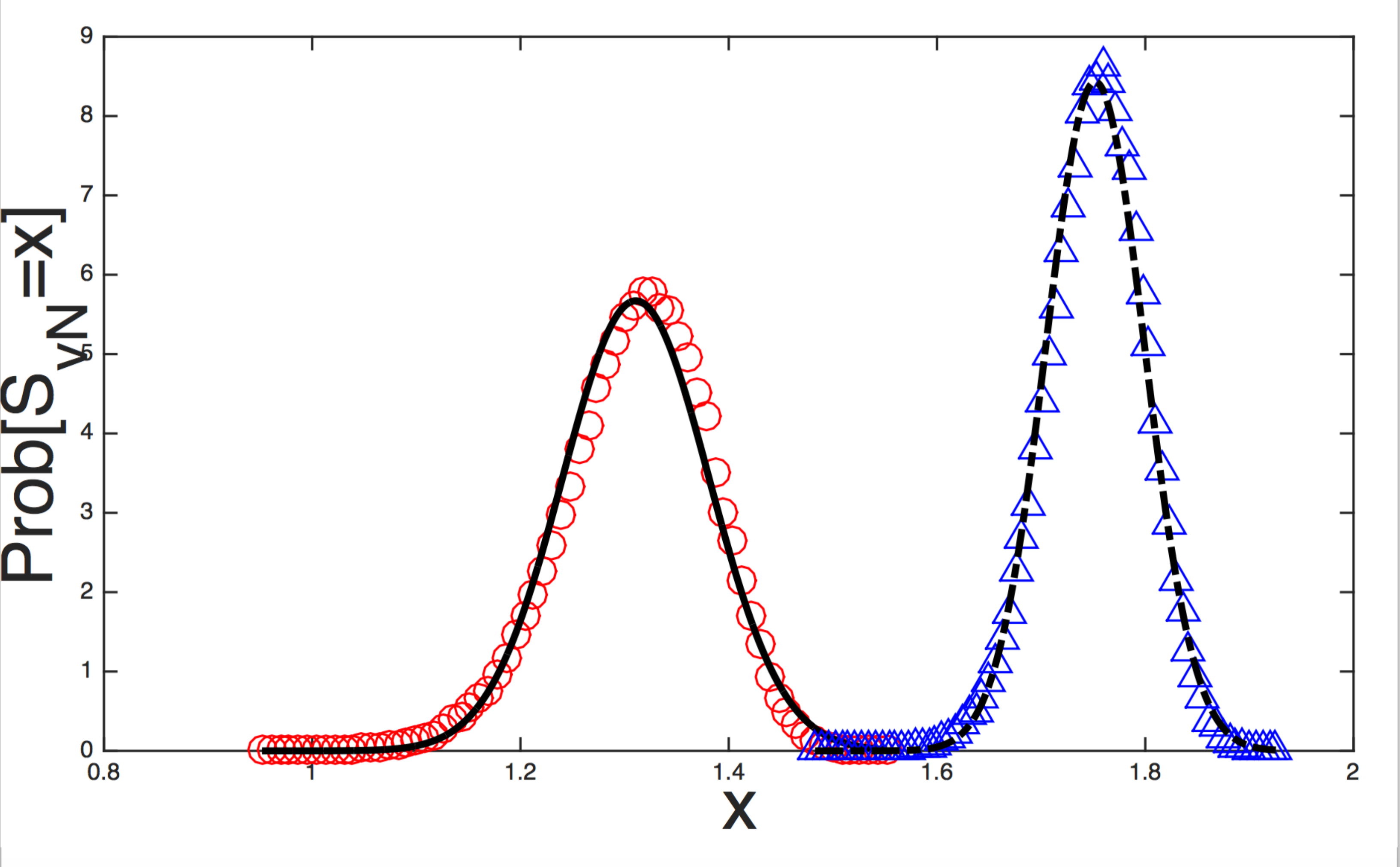}
\end{center}
\caption{(color online) Full probability density of $S_{\rm vN}$ from numerical diagonalization with $N=5,M=8$ (red circles) and $N=8,M=12$ (blue triangles), compared with Gaussian curves with mean and variance as in \eqref{mean2} and \eqref{N}, respectively.}
\label{fig4bis}
\end{figure}

In Fig.\ref{fig4} we plot the result in \eqref{N} as a function of $M$. We observe that, contrary to the behavior of
 $\langle S_{\rm vN} \rangle_{FT}$ in \eqref{mean} and \eqref{mean2}, the variance of the von Neumann entropy is a \textit{decreasing} function of $M$, which signifies that in the asymptotic limit $M \to \infty$ the distribution of the von Neumann entropy tends to a delta-function so that $S_{\rm vN}$  becomes self-averaging. In Fig. \ref{fig4bis}, we plot the full distribution of the von Neumann entropy, obtained from numerical diagonalization of fixed-trace ensembles, and compare the numerical histogram with Gaussian curves with mean and variance as in \eqref{mean2} and \eqref{N}, respectively.

\vspace{0.2in}

Finally, we turn to the asymptotic behavior of \eqref{N} 
in the limits $N = M \to \infty$, $M \to \infty$ with $N$ fixed, and the so-called double scaling limit when $N = c M$, $0 < c \leq 1$ and $M \to \infty$. For square systems with  $N = M \to \infty$ we readily find
\begin{align}
\label{squareentropyvar}
{\rm Var}\left(S_{\rm vN}\right) = \frac{1}{4 M^2} - \frac{2}{3 M^4} + \frac{14}{15 M^6} + \mathcal{O}\left(\frac{1}{M^8}\right) \,.
\end{align}
Note that the leading asymptotic 
term in this expansion coincides with the result  ${\rm Var}\left(S_{\rm vN}\right) \simeq 1/4 M^2$
obtained in \cite{majlarge,majlarge2} using a Coulomb gas method. Next, in the limit $M \to \infty$ with $N$ kept fixed and finite, we find
the following exact
 asymptotic expansion
\begin{align}
\label{asymp}
{\rm Var}\left(S_{\rm vN}\right) &= \frac{(N^2-1)}{2  N^2 M^2} \left[1 -\frac{\left(3 N^2 +10\right)}{6 N M} \nonumber\right.\\
&+\left. \frac{\left(5 N^2 +12\right)}{6 N^2 M^2}+\mathcal{O}\left(\frac{1}{N^3 M^3}\right)\right] \,.
\end{align}
Lastly, in the double scaling limit we obtain 
\begin{align}
\label{asymp2}
{\rm Var}\left(S_{\rm vN}\right) &= \frac{2 - c}{4 M^2} - \frac{(2 - c) (3 + 5 c)}{12 c^2 M^4} \nonumber\\
 &+ \frac{50 + 35 c - 27 c^2 - 2 c^3}{60 c^3 M^6} + \mathcal{O}\left(\frac{1}{M^8}\right) \,.
\end{align}
Note that in all three asymptotic expansions
the variance decreases as the second inverse power of $M$ to leading order in $M$.

\section{Schmidt number}
\label{Schmidt}

In this section, we focus on the Schmidt number $K$, defined in Eq. \eqref{K}, which is a random variable supported on $[1,N]$. Recall that it quantifies the number of effective degrees of freedom contributing to the entanglement, therefore $K = N$ corresponds to the maximally entangled state, while $K = 1$ corresponds to the completely unentangled state. 
Our formalism allows us to completely characterize the statistical properties of $K$. 

Before proceeding, it may be useful to remark that in some instances $K$ can be directly measured experimentally. In \cite{exter}, the Schmidt number has been given an operational meaning using the connection between the Schmidt decomposition in  quantum mechanics and the coherent-mode decomposition in classical coherence theory, and measured for  two-photon states entangled in a transverse mode structure.

For convenience of presentation, this rather long section is structured as follows.

In the first subsection, 
capitalizing on known results on the behavior of the purity  \cite{scott,Giraud1,Giraud2}, we will present exact formulae for the pdf of $K$ with $N = 2,3$ and arbitrary $M$, and derive the corresponding exact expressions for the moments of $K$ of arbitrary order and determine the asymptotic, leading large-$M$ behavior of the variance of $K$.

Next, we will briefly discuss the tails of the distribution in square systems with $N = M$ in the large-$N$ limit, using the results of the seminal analysis of the distribution of purity 
in \cite{majlarge,majlarge2}. 

Further on, still for square systems, we will derive exact representations for the moments of $K$ of arbitrary order expressing them via the probability $P(x_{min}^{(GUE)} \geq \sqrt{2 N} \xi)$ that 
the smallest eigenvalue $x_{min}^{(GUE)}$ of a $N \times N$ matrix belonging to the Gaussian Unitary Ensemble is greater or equal to $\sqrt{2 N} \xi$.

In the final part, we will discuss the asymptotic behavior of the moments of $K$ for square $N \times N$ systems 
in the large-$N$ limit.

\subsection{Pdf of $K$ for $N = 2$ and $N = 3$ for arbitrary $M$.}

For arbitrary $N$ and $M$, the pdf of the purity can be formally written down as
\begin{align}
&P\left(\Sigma_2\right)\!=Z^{-1}_{N,M}    \, \int \delta\left(\sum_{i=1}^N \ld_i - 1\right) \nonumber\\
&\delta\left(\Sigma_2 - \sum_{i=1}^N \ld_i^2 \right)  \, |\triangle({\bm \ld})|^{2}\, \prod_{i=1}^{N}
\ld_i^\alpha \, d\ld_i\,,
\label{jpdf11}
\end{align}
where $Z^{-1}_{N,M}$ is defined in \eqref{norm} with $\beta=2$ and $\alpha = b = M - N$. The calculation of this pdf therefore amounts to integrating the squared Vandermonde determinant, $|\triangle({\bm \ld})|^{2}$, over a domain formed by the part of the intersection of a hyperplane $\sum_i \ld_i = 1$ and a hypersphere $\sum_i \ld_i^2 = \Sigma_2$, centered at the origin, and appearing within the hypercube $[0,1]^N$.  In the general case, of course, a direct calculation of the integral in \eqref{jpdf11} for finite $N,M$ is seemingly hopeless. Giraud \cite{Giraud1,Giraud2} has nonetheless managed to overcome this challenge not only for the simple case $N = 2$, analyzed earlier by Scott and Caves \cite{scott}, but also for 
$N = 3$ and $N = 4$, which required a much more involved analysis.

\begin{figure}[ht!]
\begin{center}
\includegraphics[width=8cm]{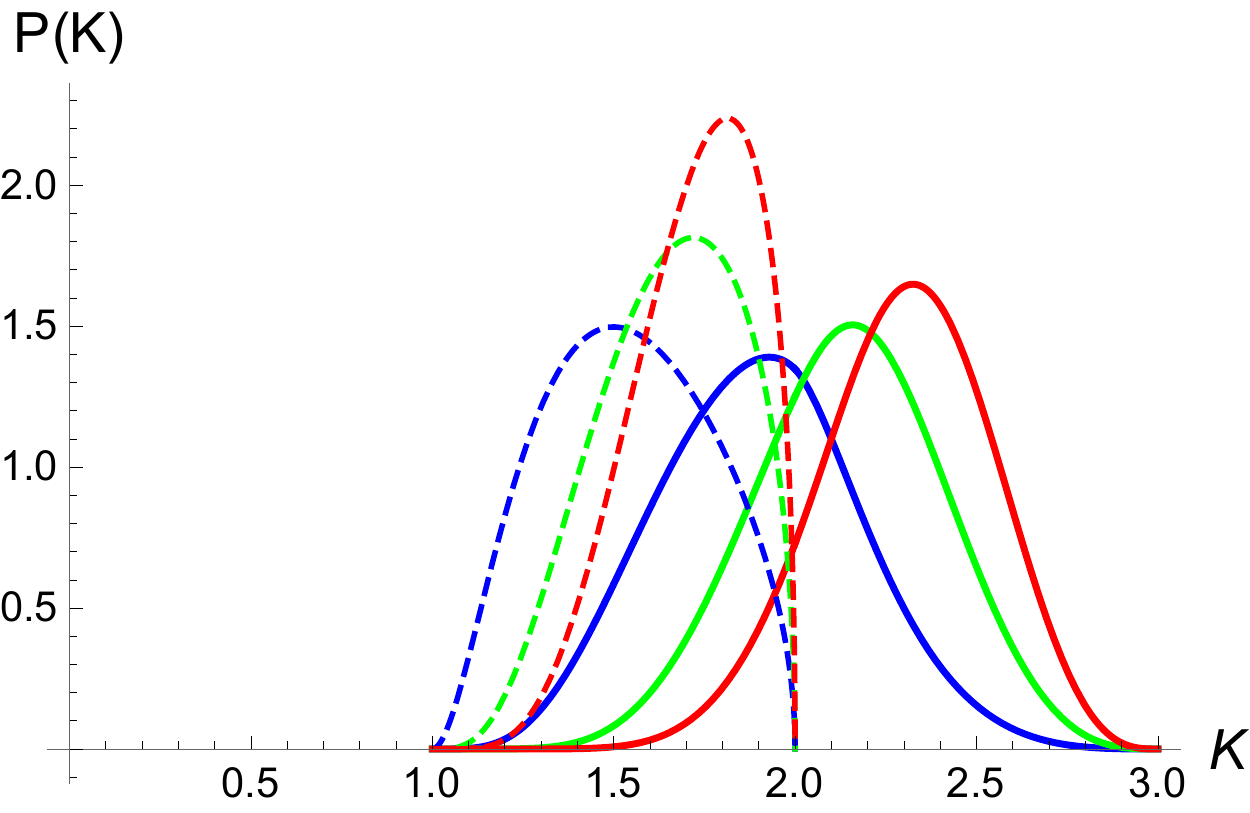}
\end{center}
\caption{(color online) Probability distribution function $P_{N=2}(K)$ for $N=2$, \eqref{KN2} (dashed lines), and $P_{N=3}(K)$ for $N=3$, \eqref{KN3} (solid lines). The curves from the left to the right correspond to $M= 4,6$ and $8$. 
}
\label{fig5}
\end{figure}

Building on the results in \cite{scott,Giraud1,Giraud2}, we can easily deduce that
for $N=2$ the pdf $P_{N=2}(K)$ of the Schmidt number $K$ is given by
\begin{equation}
\label{KN2}
P_{N=2}(K)=\frac{2^M \Gamma(M+1/2)}{\sqrt{\pi} \Gamma(M-1)} \frac{\left(K-1\right)^{M-2}\sqrt{2-K}}{K^{M+1/2}}
\end{equation}
for $K \in [1,2]$ and is zero otherwise. Further on, for $N = 3$ an analogous result is given by a more complicated formula 
\begin{align}
\label{KN3}
P_{N=3}(K)&= \dfrac{(3 M - 1)! }{16 \sqrt{3} \prod_{j=1}^3 (M-j)!  K^2} \left(\dfrac{5 K - 9}{54 K}\right)^{M-3} \nonumber\\
&\times \sum_{k=0}^{M-3} (-1)^k \binom{M-3}{k} \left(\dfrac{9 K}{\sqrt{6} (5 K - 9)}\right)^k \nonumber\\
&\times \left(\dfrac{3 - K}{3 K}\right)^{3 (k+2)/2} \sum_{j=0}^{\lfloor k/2 \rfloor} \left(1 - \delta_{j,0} \delta_{\bar{k},0}\right) \nonumber\\
&\times  \binom{k}{\lfloor k/2 \rfloor - j} \left(\zeta_{j+\bar{k}/2}(\phi) - \zeta_{j+\bar{k}/2}(\pi/3)\right) \,, \nonumber\\
\end{align}
where $\lfloor k/2 \rfloor$ is the floor function, 
$\bar{k} = k\mbox{ mod }2$, $\delta_{k,j}$ is the Kronecker delta symbol, 
$\phi = 0$ for $K \in [2,3] $ and $\phi = \arccos(\sqrt{K/(6 - 2 K)})$ for $K \in [1,2]$ (so that $P_{N=3}(K)$ is piece-wise continuous). In Fig. \ref{fig5} we plot the distribution function in \eqref{KN2} and \eqref{KN3}.

\vspace{0.2in}

The behavior of the pdf of the purity for square systems with $N = M$ has been analyzed in the  large-$N$ limit in
\cite{majlarge,majlarge2} using a Coulomb gas method. It was there shown that the pdf exhibits three different regimes, associated with a different behavior of the equilibrium density of eigenvalues: a left tail, a central Gaussian peak, and a right tail, similarly to what we observe in Fig. \ref{fig5}. Translating these results in the language of Schmidt number, the pdf $P(K)$ in the vicinity of the left edge of the support, i.e., for $K$ close to $1$, should tend to zero as
\begin{equation}
P(K) \sim \left(1 - 1/\sqrt{K}\right)^{N^2}/K^2 \ .
\end{equation}
In the opposite limit, i.e. when $K$ is close to $N$ so that the system tends to a maximally entangled state, 
\begin{equation}
P(K) \sim \left(\dfrac{N}{K}-1\right)^{N^2/2}/K^2\ .
\end{equation}
These two asymptotic regimes are separated by a Gaussian peak near $N/2$.

\subsection{Moments of $K$ for $N=2,3$ and arbitrary $M$, and moments of purity}

For $N=2$ the moments of $K$ of arbitrary (not necessarily integer) order $a$ 
can be readily found from \eqref{KN2} and
read
\begin{equation}
\mathbb{E}_{FT}\left\{ K^{a}_{N=2}\right\} = \,_2F_1\left(M-1,a;M+1/2;1/2\right) \,,
\end{equation}
where $_2F_1(\ldots)$ is the Gauss hypergeometric function. Hence, for sufficiently large $M$, 
$\mathbb{E}_{FT}\left\{ K^{a}_{N=2}\right\}$ is conveniently represented by (see \eqref{asympF} in the Appendix \ref{appendix3})
\begin{equation}
\label{momN=2}
\mathbb{E}_{FT}\left\{ K^{a}_{N=2}\right\} = 2^{a} \Big(1 - \dfrac{3 a}{2 M} +  \dfrac{3 a (7 + 5 a)}{8 M^2} + \mathcal{O}\left(\dfrac{1}{M^3}\right) \Big)\,,
\end{equation}
so that, to leading order in $M$, the variance of $K_{N=2}$ is given by
\begin{equation}
{\rm Var}\left(K_{N=2}\right) = \dfrac{6}{M^2} + \mathcal{O}\left(\dfrac{1}{M^3}\right) \,.
\end{equation}
For $N=3$, the calculations are rather involved and we relegate them to the Appendix \ref{appendix3}, 
where we derive an explicit, albeit rather cumbersome
 exact result for $\mathbb{E}_{FT}\left\{ K^{a}_{N=3}\right\}$ (see \eqref{formal3} in  the Appendix \ref{appendix3}). 
The asymptotic large-$M$ behavior 
turns out to have a rather simple form and for 
arbitrary, not necessarily integer $a$, we have
\begin{align}
\label{momN=3}
&\mathbb{E}_{FT}\left\{ K^{a}_{N=3}\right\}  = 3^{a} \Big(1 - \dfrac{7 \, a}{3\, M} + \nonumber\\
& + \dfrac{7 \, a (9 a + 11)}{18 \, M^2} + \mathcal{O}\left(\dfrac{1}{M^3}\right) \Big) \,,
\end{align}
so that the variance of $K_{N=3}$, to leading order in $M$, obeys
\begin{equation}
\label{varN3}
{\rm Var}\left(K_{N=3}\right) = \dfrac{14}{M^2} + \mathcal{O}\left(\dfrac{1}{M^3}\right) \,.
\end{equation}

\begin{figure}[ht!]
\begin{center}
\includegraphics[width=8cm]{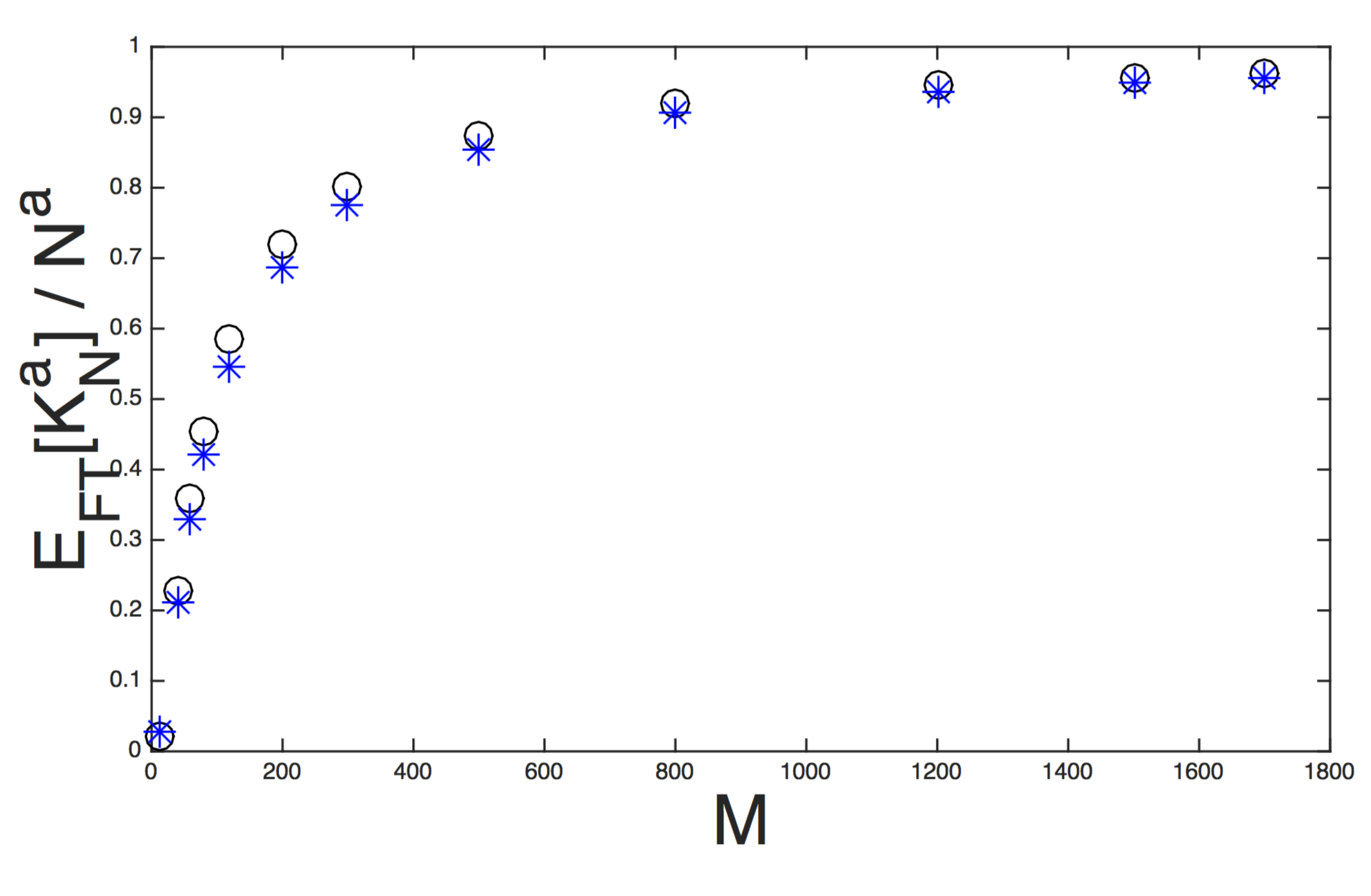}
\end{center}
\caption{(color online) $\mathbb{E}_{FT}\left\{ K^{a}_{N}\right\}/N^a$ as a function of $M$, for $N=11,a=6.2$ (black circles) and $N=25,a=3.2$ (blue stars). The convergence to the theoretical value $1$ is quite convincing. 
}
\label{figM}
\end{figure}

We observe that the leading terms in \eqref{momN=2} and \eqref{momN=3} are $2^{a}$ (for $N=2$) and $3^{a}$ (for $N=3$). It seems natural to conjecture that for general $N$ the leading term should be $N^a$. 
While we are unable to prove 
 that
\begin{equation}
\label{z8}
\lim_{M \to \infty} \mathbb{E}_{FT}\left\{ K^{a}_{N}\right\} = N^{a} \ , 
\end{equation}
such an assumption seems to be quite plausible on physical grounds and
signifies that in situations in which a small subsystem of size $N$ is attached to a much larger subsystem of size $M$, a complete entanglement is achieved when $M \to \infty$. This large-$M$ behavior is corroborated by numerical simulations, shown in Fig. \ref{figM}.

\vspace{0.2in}

To close this subsection, we take advantage of the
works by Scott and Caves \cite{scott} and Giraud \cite{Giraud2}, who calculated exactly the first few cumulants of the purity $\Sigma_2$, 
and extract from these results the asymptotic, large-$M$ behavior of the moments of $\Sigma_2$ with fixed $N$. We find that, explicitly,
\begin{equation}
\mathbb{E}_{FT}\left\{\Sigma_2\right\}  = \dfrac{1}{N} \left(1 + \dfrac{N^2 - 1}{N \, M} - \dfrac{N^2 - 1}{N^2 \, M^2}
+ \mathcal{O}\left(\dfrac{1}{M^3}\right)\right)\ ,
\end{equation}
\begin{align}
&\mathbb{E}_{FT}\left\{\Sigma_2^2\right\} = \dfrac{1}{N^2} \Big(1 + \dfrac{2 \left(N^2 - 1\right)}{N \, M} + \nonumber\\
&+ \dfrac{\left(N^2 - 1\right)^2}{N^2 \, M^2}
+ \mathcal{O}\left(\dfrac{1}{M^3}\right)\Big)\ ,
\end{align}
\begin{align}
&\mathbb{E}_{FT}\left\{\Sigma_2^3\right\} = \dfrac{1}{N^3} \Big(1 + \dfrac{3 \left(N^2 - 1\right)}{N \, M} + \nonumber\\
&+ \dfrac{3 \left(N^2 - 1\right)}{M^2}
+ \mathcal{O}\left(\dfrac{1}{M^3}\right)\Big)\ ,
\end{align}
and
\begin{align}
&\mathbb{E}_{FT}\left\{\Sigma_2^4\right\} = \dfrac{1}{N^4} \Big(1 + \dfrac{4 \left(N^2 - 1\right)}{N \, M} + \nonumber\\&+\dfrac{6 N^4 - 4 N^2 - 2}{N^2 \, M^2}
+ \mathcal{O}\left(\dfrac{1}{M^3}\right)\Big)\ ,
\end{align}
while the variance of $\Sigma_2$, which is given exactly by \cite{scott,Giraud1,Giraud2}
\begin{align}
\label{varsigmapurity}
{\rm Var}\left(\Sigma_2\right) = \dfrac{2 \left(N^2 - 1\right) \left(M^2 - 1\right)}{\left(1 + N \, M\right)^2 \, \left(2 + N \, M\right) \, \left(3 + N \, M\right)}
\end{align}
obeys, to leading order in $M$,
\begin{align}
\label{varsigmapur}
{\rm Var}\left(\Sigma_2\right) = \dfrac{2 \left(N^2 - 1\right)}{N^4 \, M^2} + \mathcal{O}\left(\dfrac{1}{M^3}\right)\ .
\end{align}
We note that as $M \to \infty$ and at a fixed $N$, the purity tends to the value $1/N$, i.e., attains the lower limit of its support. Together with our \eqref{z8}, it implies that systems with fixed $N$ and $M \to \infty$ become completely entangled.

\subsection{Moments of $K$ for square $N = M$ systems: general expressions}

Having discussed the properties of $\mathbb{E}_{FT}\left\{ K^{a}\right\}$ for fixed $N$ and varying $M$, 
we turn to the analysis of the moments of $K$ in the special quadratic 
case $N=M$ (so that $\alpha = b =0$ in \eqref{jpdf1}). 
The general case when $N \leq M$ can also be considered within the framework we present below 
but will require much lengthier calculations.
We make use of
our \eqref{bb}, set $\beta = 2$, $n=N$  and put $p_i = 1/(8 N\xi^2)$ with $\xi \geq 0$. 
Recalling next the definition of $K$, we
rewrite \eqref{bb} as
\begin{align}
\label{bbb3}
(2 N)^{N^2/2} &\xi^{N^2} \mathbb{E}_{FT}\left\{ K^{N^2/2} U\left(\frac{N^2}{2},\frac{1}{2}, 2 N \xi^2 K\right)\right\}\nonumber\\ 
 & =  \mathbb{E}_{WL}\left\{ \exp\left( - \frac{1}{32 N \xi^2} \sum_{i=1}^{N} y_i^2\right) 
 \right\} \,.
\end{align}
Using \eqref{jpdf2}, the right-hand-side of \eqref{bbb3} can be written explicitly as 
\begin{align}
\dfrac{Z^{-1}_{N,N}}{2^{N^2} \Gamma(N^2)} \int_0^{\infty} \triangle^2({\bm y})\, \prod_{i=1}^{N} \exp\left( - \frac{y_i}{2} - \frac{y_i^2}{32 N \xi^2} \right) dy_i \ , \nonumber\\
\end{align}
where $Z_{N,N}$ is defined in \eqref{norm} with $\beta$ set equal to $2$ and $N = M$. Making a linear shift
of the integration variables $y_i \to 4 \sqrt{2 N} \xi (x_i - \sqrt{2 N} \xi)$, we formally rewrite the latter expression as
\begin{align}
\label{guen}
&\dfrac{Z^{-1}_{N,N} (8 N)^{N^2/2}}{\Gamma(N^2)} \xi^{N^2} \,  \exp\left(2 N^2 \xi^2\right)  \nonumber\\
&\times\int_{\sqrt{2 N} \xi}^{\infty} \triangle^2({\bm x})\, \prod_{i=1}^{N} \exp\left( - x^2_i \right) dx_i \,.
\end{align}
One notices next that the integral in the second line in \eqref{guen}, up to the normalisation factor $Z^{-1}_{N,N}(GUE)$,
\begin{equation}
\label{normGUE}
Z^{-1}_{N,N}(GUE) = \dfrac{2^{N (N-1)/2}}{\pi^{N/2} \prod_{j=1}^N \Gamma(j+1)} \,,
\end{equation}
is equal to the probability $P\left(x_{min}^{(GUE)} \geq \sqrt{2 N} \xi\right)$ that 
the smallest eigenvalue $x_{min}^{(GUE)}$ in a $N \times N$ Gaussian Unitary Ensemble is greater or equal to $\sqrt{2 N}\xi$ \cite{mehta,widom}.
Consequently, combining \eqref{bbb3} and \eqref{guen}, we establish the following intriguing representation 
\begin{align}
\label{b4}
&\mathbb{E}_{FT}\left\{ K^{N^2/2} U\left(\frac{N^2}{2},\frac{1}{2}, 2 N \xi^2 K\right)\right\} = \frac{(2 \pi)^{N/2} 2^{N^2/2}}{G(N+1)} \nonumber\\ 
&\times \exp\left(2 N^2 \xi^2\right) P\left(x_{min}^{(GUE)} \geq \sqrt{2 N} \xi\right) \,, 
\end{align}
where $G(N+1) = \prod_{j=0}^{N-1} j!$ is the Barnes G-function. This result holds for arbitrary $N$ and arbitrary non-negative $\xi$ and will be used in what follows for the derivation of explicit and exact representations of $\mathbb{E}_{FT}\left\{ K^{a}\right\}$.

\subsubsection{Moments of $K$ of order $0<a < N^2/2$}

Moments of order $a < N^2/2$ can be obtained as follows: multiplying both sides of \eqref{b4} by $\xi^{N^2-2 a - 1}$, where $0 < a < N^2/2$, 
and integrating from zero to infinity, we get
\begin{align}
\label{expressionK}
&\mathbb{E}_{FT}\left\{ K^{a}\right\} = \frac{2 (2 \pi)^{N/2} (4 N)^{N^2/2}}{(8 N)^{a} G(N+1)} \frac{\Gamma(N^2)}{\Gamma(a) \Gamma(N^2- 2 a)} \nonumber\\
&\times \int^{\infty}_0 d\xi \, \xi^{N^2 - 2 a -1} \, \exp\left(2 N^2 \xi^2\right) \, P\left(x_{min}^{(GUE)} \geq \sqrt{2 N} \xi\right) \,, \nonumber\\
\end{align} 
which defines the positive 
moments of the Schmidt number of order $a < N^2/2$. Note that this constraint on the order of moments stems from the restriction on the convergence of the integral in \eqref{expressionK}; for $a \geq N^2/2$ the integral in the latter equation becomes divergent on the upper integration limit so that we have to resort to a different strategy.
In Appendix \ref{appendixC} 
we present an alternative derivation of
the result in \eqref{expressionK} using a different line of thought, closer to the derivation of the relation between the $n$-point densities of the FT and the WL ensembles.

\subsubsection{Moments of $K$ of order $a = N^2/2$}

The moment of order $a = N^2/2$ can be readily obtained from \eqref{b4} by noticing that
\begin{equation}
\lim_{\xi \to 0} U\left(\frac{N^2}{2},\frac{1}{2}, 2 N \xi^2 K\right) = \dfrac{\sqrt{\pi}}{\Gamma\left(\dfrac{N^2+1}{2}\right)} \,,
\end{equation}
so that, for arbitrary $N$, we have
\begin{align}
\label{a=N^2/2}
\mathbb{E}_{FT}\left\{ K^{N^2/2}\right\} &= \dfrac{2 \, (2 \pi)^{N/2} \Gamma\left(N^2\right)}{2^{N^2/2} \Gamma\left(N^2/2\right) G(N+1)}  \nonumber\\ &\times P\left(x_{min}^{(GUE)} \geq 0\right) \,.
\end{align}
This expression relates, at a first glance rather surprisingly, the moment of $K$ of order $N^2/2$ in the FT ensemble to the probability of the highly atypical event 
where all the eigenvalues of a $N \times N$ matrix belonging to the GUE are positive. 

\subsubsection{Moments of $K$ of order $a > N^2/2$}

The moments of order $N^2/2 + m$ or $N^2+ 1/2 + m$, $m$ being an integer,
can be accessed in the following way. From \eqref{b4}, recall the Taylor series expansion of the Tricomi's confluent hypergeometric function and express  
$P\left(x_{min}^{(GUE)} \geq \sqrt{2 N} \xi\right)$ with non-negative $\xi$ in terms of the moments of $K$, which gives 
\begin{align}
\label{sea}
&P\left(x_{min}^{(GUE)} \geq \sqrt{2 N} \xi\right) = \frac{\sqrt{\pi} 2^{N^2/2} G\left(N+1\right)}{2 (2 \pi)^{N/2} \Gamma\left(N^2\right)} e^{- 2 N^2 \xi^2} \nonumber\\
& \Big(\sum_{m=0}^{\infty} \frac{\Gamma\left(\frac{N^2}{2}+ m\right)}{\Gamma\left(m+\frac{1}{2}\right)} \mathbb{E}_{FT}\left\{K^{\frac{N^2}{2} + m}\right\}  \frac{\left(2 N \xi^2\right)^m}{m!}  \nonumber\\
&-  \sum_{m=0}^{\infty} \frac{\Gamma\left(\frac{N^2+1}{2}+ m\right)}{\Gamma\left(m+\frac{3}{2}\right)} \mathbb{E}_{FT}\left\{K^{\frac{N^2+1}{2} + m}\right\}  \frac{\left(2 N \xi^2\right)^{m+1/2}}{m!}\Big) \,. \nonumber\\
\end{align}
The latter equation states that $P\left(x_{min}^{(GUE)} \geq \sqrt{2 N} \xi\right)$ can be interpreted as
 the generating function of moments of $K$ of order higher than $N^2/2$. In other words, \eqref{sea} is the Taylor series expansion of $\exp{(2 N^2 \xi^2)} P\left(x_{min}^{(GUE)} \geq \sqrt{2 N} \xi\right)$ in powers of $\xi^2$ and the coefficients in this expansion are just the derivatives of the latter function at $\xi = 0$:
 
\begin{align}
\label{moments2}
& \mathbb{E}_{FT}\left\{ K^{N^2/2+m} \right\} = \frac{2 (2 \pi)^{N/2} \Gamma(N^2)}{2^{N^2/2} \Gamma(N^2/2+m) G(N+1)} \nonumber\\
&\left. \left( \dfrac{d^{2m}}{d \xi^{2m}} \exp\left(2 N^2 \xi^2\right) \, P\left(x_{min}^{(GUE)} \geq \sqrt{2 N} \xi\right) \right)\right|_{\xi=0}
\end{align}
and
\begin{align}
&\mathbb{E}_{FT}\left\{ K^{N^2/2+m+1/2} \right\} = - \frac{(2 \pi)^{N/2}}{\sqrt{2} \, 2^{3m + N^2/2} N^{m + 1/2} } \nonumber\\
&\dfrac{ \Gamma(N^2)}{\Gamma(N^2/2+m+1/2) G(N+1)} \nonumber\\
&\left. \left( \frac{d^{2m+1}}{d \xi^{2m+1}} \exp\left(2 N^2 \xi^2\right) \, P\left(x_{min}^{(GUE)} \geq \sqrt{2 N} \xi\right) \right)\right|_{\xi=0}\,.
\end{align}
Therefore, the moments of the Schmidt number $K$ of order greater than $N^2/2$ probe the behavior of $P\left(x_{min}^{(GUE)} \geq \sqrt{2 N} \xi\right)$ right in the middle of the Wigner sea. Since $x_{min}^{(GUE)}$ is typically $\sim - \sqrt{2 N}$, i.e., $\xi \sim - 1$, we conclude that 
the behavior of the moments of $K$ of such a high order is supported by atypical, rare events.

\subsubsection{Moments of $K$ of order $a < 0$}

Lastly, we consider the negative moments of $K$, i.e., positive moments of the purity $\Sigma_2$. 
We recall that formally exact expressions for $\mathbb{E}_{FT}\{\Sigma_2^n\}$ 
have already been computed by Giraud (see \cite{Giraud1} and some corrections in \cite{Giraud2}). Our goal here is to provide an alternative derivation, relating $\mathbb{E}_{FT}\{\Sigma_2^n\}$ to the behavior of 
$P\left(x_{min}^{(GUE)} \geq \sqrt{2 N} \xi\right)$ in the limit $\xi \to \infty$. This derivation will also allow us   to establish an exact asymptotic expansion for $P\left(x_{min}^{(GUE)} \geq \sqrt{2 N} \xi\right)$ for finite $N$.
Setting in \eqref{kkk} all $p_i = p \geq 0$ and $n$ equal to $N$, we have that for any integer $n \geq 0$,
\begin{align}
\mathbb{E}_{FT}\left\{ \Sigma_2^{n} \right\} &=  \frac{\Gamma\left(N^2\right)}{4^{n} \Gamma\left(N^2 + 2 n\right)}   \mathbb{E}_{WL}\left\{ \left(\sum_{i=1}^{N} y_i^2\right)^{n}\right\} \,.
\end{align} 
Further on, using the relation
\begin{equation}
\left(\sum_{i=1}^{N} y_i^2\right)^{n} = (-1)^{n} (32 N)^{n} \left. \frac{d^{n}}{d p^{n}} \exp\left(- \frac{p}{32 N} \sum_{i=1}^{N} y_i^2\right) \right|_{p=0} \,,
\end{equation}
we find, after some straightforward manipulations, the following expression 
\begin{align}
\label{LLL2}
&\mathbb{E}_{FT}\{ \Sigma_2^{n} \} = \dfrac{(-1)^{n} \left(8 N\right)^{n} }{ \Gamma\left(N^2 + 2 n\right)} \dfrac{ (2 \pi)^{N/2} \left(4 N\right)^{N^2/2} \Gamma\left(N^2\right)}{G(N+1)}\nonumber\\
\times & \left. \left(\dfrac{d^{n}}{dp^{n}} \dfrac{\exp\left(\dfrac{2 N^2}{p}\right)}{p^{N^2/2}} P\left(x_{min}^{(GUE)} \geq \sqrt{\dfrac{2 N}{p}}\right)\right)\right|_{p=0} \,. \nonumber\\
\end{align}
The latter expression shows that the
moments of the purity, or the
 inverse moments of $K$, are dominated by the right tail of the probability 
$P\left(x_{min}^{(GUE)} \geq \sqrt{2 N} \xi\right)$ corresponding to the limit $\xi \to \infty$.

As a certain by-product of the above considerations, which will be also useful for our further analysis, 
we obtain from \eqref{LLL2} the following \emph{exact} asymptotic representation of the probability 
$P\left(x_{min}^{(GUE)} \geq \sqrt{2 N} \xi\right)$ in the limit $\xi \to \infty$
\begin{align}
\label{asy}
&P\left(x_{min}^{(GUE)} \geq \sqrt{2 N} \xi\right) = \dfrac{G\left(N+1\right)}{(2 \pi)^{N/2} (4 N)^{N^2/2} \Gamma\left(N^2\right)} \,  \nonumber\\ 
&\times \xi^{-N^2} \exp\left(- 2 N^2 \xi^2\right) \nonumber\\
&\times
 \sum_{n=0}^{\infty} \dfrac{(-1)^{n}}{n!} \dfrac{\Gamma\left(N^2 + 2 n\right)}{(8 N \xi^2)^{n}} \mathbb{E}_{FT}\left\{\Sigma_2^{n}\right\} \,,
\end{align}
which is valid for arbitrary $N$.  Note that, in contrast to the result in \eqref{sea}, which shows that the coefficients in the expansion of 
$P\left(x_{min}^{(GUE)} \geq \sqrt{2 N} \xi\right)$ right 
in the middle of the Wigner sea are related to the moments of the Schmidt number of order $N^2/2 +m$,  the coefficients in the Taylor series for $\xi \to \infty$ are proportional to the inverse moments of $K$, i.e., the positive moments of the purity $\Sigma_2$. Therefore, $P\left(x_{min}^{(GUE)} \geq \sqrt{2 N} \xi\right)$ in the limit $\xi \to \infty$ can be thought of as the moment-generating function of the purity.

Next, capitalizing on the exact results for the moments of purity calculated by Giraud \cite{Giraud1,Giraud2}, 
we rewrite the exact asymptotic large-$\xi$ expansion in \eqref{asy} in a closed form:
\begin{align}
\label{bbb}
&P\left(x_{min}^{(GUE)} \geq \sqrt{2 N} \xi\right) = \dfrac{G\left(N+1\right) \Gamma\left(N+1\right)}{(2 \pi)^{N/2} (4 N)^{N^2/2}} \,  \nonumber\\ 
&\times \xi^{-N^2} \exp\left(- 2 N^2 \xi^2\right)
 \sum_{n=0}^{\infty} (-1)^{n} \dfrac{B_n}{\left(8 N \xi^2\right)^n}  \,,
 \end{align}
 where
 \begin{align}
&B_n = \sum_{n_1+n_2+ \ldots + n_N = n} \Big(\prod_{i=1}^N \dfrac{\left(N+ 2 n_i - i\right)!}{i! n_i! (N - i)!} \nonumber\\
&\times \prod_{1 \leq i < j \leq N} \left(2 n_i - i - 2 n_j + j\right) \Big) \,,
\end{align}
which is a finite sum over partitions of $n$ into $N$
numbers greater or equal to 0. These partitions can
be easily generated for any $n$ by some suitable algorithm. In particular, the first four terms in the expansion in \eqref{bbb} (or, equivalently, in \eqref{asy}) are given explicitly by
\begin{align}
\label{bb4}
&P\left(x_{min}^{(GUE)} \geq \sqrt{2 N} \xi\right) = \dfrac{G\left(N+1\right)}{(2 \pi)^{N/2} (4 N)^{N^2/2}} \,  \nonumber\\ 
&\times \xi^{-N^2} \exp\left(- 2 N^2 \xi^2\right)
\Big(1 - \dfrac{N^2}{4 \, \xi^2} + \dfrac{\left(2 N^4 + 9 N^2 +1\right) }{64 \, \xi^4} - \nonumber\\
&- \dfrac{\left(2 N^6 + 27 N^4 +111 N^2 + 40   \right)}{768 \, \xi^6} + \mathcal{O}\left(\dfrac{1}{\xi^8}\right)
\Big) \,.
 \end{align}
 To the best of our knowledge, the result in \eqref{bbb} is new.  

\subsection{Moments of $K$ and of the purity $\Sigma_2$ for square $N \times N$ systems: asymptotic large-$N$ behavior}

In this final section we   
focus on the asymptotic large-$N$ behavior of the moments of $K$ of order $a < N^2/2$
and of the purity $\Sigma_2$ for square, $N \times N$ systems, a question that seemingly 
has not been addressed as yet.
The large-$N$ behavior of the moments 
of $\Sigma_2$ will be simply extracted from the available exact expressions for the first few cumulants of the purity\cite{scott,Giraud1,Giraud2}. On the other hand, to deduce the large-$N$ behavior of the moments of $K$ we will resort to the classical papers by Dean and Majumdar \cite{DM}, who have shown 
that for sufficiently large $N$ the probability 
$P\left(x_{min}^{(GUE)} \geq \sqrt{2 N} \xi\right)$ has the form
\begin{equation}
\label{David}
P(x_{min}^{(GUE)} \geq \sqrt{2 N} \xi) \approx  \exp\left(- 2 N^2 \Phi\left(\xi\right) + \mathcal{O}\left(N\right)\right)
\end{equation}
where the symbol $"\approx"$ means equivalence on a logarithmic scale, and
the large deviation function $\Phi(\xi)$  in \eqref{David}  is given explicitly by \cite{DM}:
\begin{align}
\label{Satya}
&\Phi\left(\xi\right) = \frac{1}{2} \ln3 - \frac{1}{2} \ln\left(\sqrt{3 + \xi^2} - \xi\right) \nonumber\\
&+\frac{1}{54} \left(36 \xi^2- 2 \xi^4 + \xi \left(15 + 2 \xi^2\right) \sqrt{3 + \xi^2}\right) \,.
\end{align} 

Note that, 
as evidenced by numerical simulations performed in \cite{DM}, the result in \eqref{David} and \eqref{Satya} is reasonably accurate already for quite modest values of $N$.

Consequently, in this subsection 
we will posit that the moments of $K$ of order less than $N^2/2$, defined
by our \eqref{expressionK}, can be approximated with logarithmic accuracy by
\begin{align}
\label{approximationK}
&\mathbb{E}_{FT}\left\{ K^{a}\right\} \approx \frac{2 (2 \pi)^{N/2} (4 N)^{N^2/2}}{(8 N)^{a} G(N+1)} \frac{\Gamma(N^2)}{\Gamma(a) \Gamma(N^2- 2 a)}   \nonumber\\
&\times C_N \, \int^{\infty}_0 d\xi \, \xi^{N^2 - 2 a - 1} \, \exp\left(2 N^2  \xi^2\right)  \exp\left(- 2 N^2 \Phi\left(\xi\right)\right) \,,
\end{align}
where the constant $C_N$ will be defined below.

Before we proceed with the analysis of the
expression in \eqref{approximationK},
we find it  necessary to emphasize several points on the possible limitations of the result in 
\eqref{David} and \eqref{Satya}, and consequently, on the errors these limitations may incur in determining the asymptotical behavior of the moments of the Schmidt number for square systems.

(i) Note that due to the omitted $\mathcal{O}(N)$ terms in \eqref{Satya}, the limiting value $\lim_{a \to 0} \mathbb{E}_{FT}\{K^{a}\}$ in \eqref{approximationK} is not equal to $1$, as it should. We introduce \emph{ad hoc} in \eqref{approximationK}  a normalization constant $C_N$, which will \emph{enforce} the condition $\lim_{a \to 0} E_{FT}\{K^{a}\} \equiv 1$. Clearly, this normalization constant should not dependent on $N$ stronger then $\exp\left(\mathcal{O}(N)\right)$. We present the calculation of this constant in Appendix \ref{appendixD} and show that $C_N$ is given explicitly by 
\begin{equation}
\label{CNn}
C_N = \dfrac{G(N+1)}{(2 \pi)^{N/2} N^{N^2/2}} \exp\left(\dfrac{3 N^2}{4}\right) \,.
\end{equation}
We observe that, indeed, this $C_N$ depends very weakly on $N$ (see the Appendix \ref{appendixD}).

(ii)  Let us examine next the large-$\xi$ behavior of 
the large deviation form $P(x_{min}^{(GUE)} \geq \sqrt{2 N} \xi)$  in \eqref{David} and \eqref{Satya} and compare
 it with our exact asymptotic expansion in \eqref{bb4}.  The  large-$\xi$ asymptotic of the expression in \eqref{David} and \eqref{Satya} can be readily determined and reads
\begin{align}
\label{largexi}
&P(x_{min}^{(GUE)} \geq \sqrt{2 N} \xi) \approx \exp\left(- 2 N^2 \Phi\left(\xi\right)\right) = \nonumber\\
&=\dfrac{\exp\left(- \dfrac{3 N^2}{4}\right)}{2^{N^2}} \xi^{-N^2} \exp\left( - 2 N^2 \xi\right) \times \nonumber\\
&\times \Big(1 - \frac{N^2}{4 \, \xi^4} + \frac{\left(2 N^4 + 9 N^2\right)}{64 \, \xi^4} - \nonumber\\
&- \frac{\left(2 N^6 + 27 N^4 + 108 N^2\right)}{768 \, \xi^6} + \mathcal{O}\left(\dfrac{1}{\xi^8}\right)\Big) \,.
\end{align}
Further on, the $N$-dependent 
numerical factor in the first line in \eqref{bb4} 
can be written down, for sufficiently large $N$,  as
\begin{align}
\label{ququ}
& \dfrac{G\left(N+1\right)}{(2 \pi)^{N/2} (4 N)^{N^2/2}} \sim \dfrac{\exp\left(- \dfrac{3 N^2}{4}\right)}{2^{N^2}} \times \nonumber\\
 &\times \dfrac{1}{A} \left(\dfrac{e}{N}\right)^{1/12} \exp\left(\mathcal{O}\left(\dfrac{1}{N^2}\right)\right) \,,
\end{align}
where $A \approx 1.282...$ is the Glaisher constant 
and $e$ is the base of the natural logarithm. 
Comparing next both expansions, 
we first observe that
 the 
 leading terms coincide, once we discard the correction terms in the second line in \eqref{ququ}. In fact, the constant $C_N$ in \eqref{CNn}, chosen to enforce the condition $\lim_{a \to 0} \mathbb{E}_{FT}\{K^{a}\} \equiv 1$, is actually equal to these discarded correction terms.  
 Next, the subdominant terms in both expansion are series in powers of $\xi^{-2}$ (multiplied by $\xi^{-N^2} \exp(- 2 N^2 \xi)$) and the coefficients in this  series are polynomials of $N$.
Inspecting the expansion coefficients in \eqref{bb4} and \eqref{largexi}, we note that the first two terms coincide exactly, while 
the third and the fourth ones are only slightly different, which means that the omitted $\mathcal{O}(N)$ terms start to contribute only at this level. Concluding this discussion, we may expect that $P(x_{min}^{(GUE)} \geq \sqrt{2 N} \xi)$ in \eqref{David} defines correctly the leading and the first sub-leading terms in the moments of purity. On the other hand, since the moments of $K$ are defined as an integral over $\xi$ and we do not know how accurately  $P(x_{min}^{(GUE)} \geq \sqrt{2 N} \xi)$ in \eqref{David} approximates the true $P(x_{min}^{(GUE)} \geq \sqrt{2 N} \xi)$  for small and moderate $\xi$, we can not expect the same level of accuracy.

\begin{figure}[!ht]
\begin{center}
\includegraphics[width=8cm]{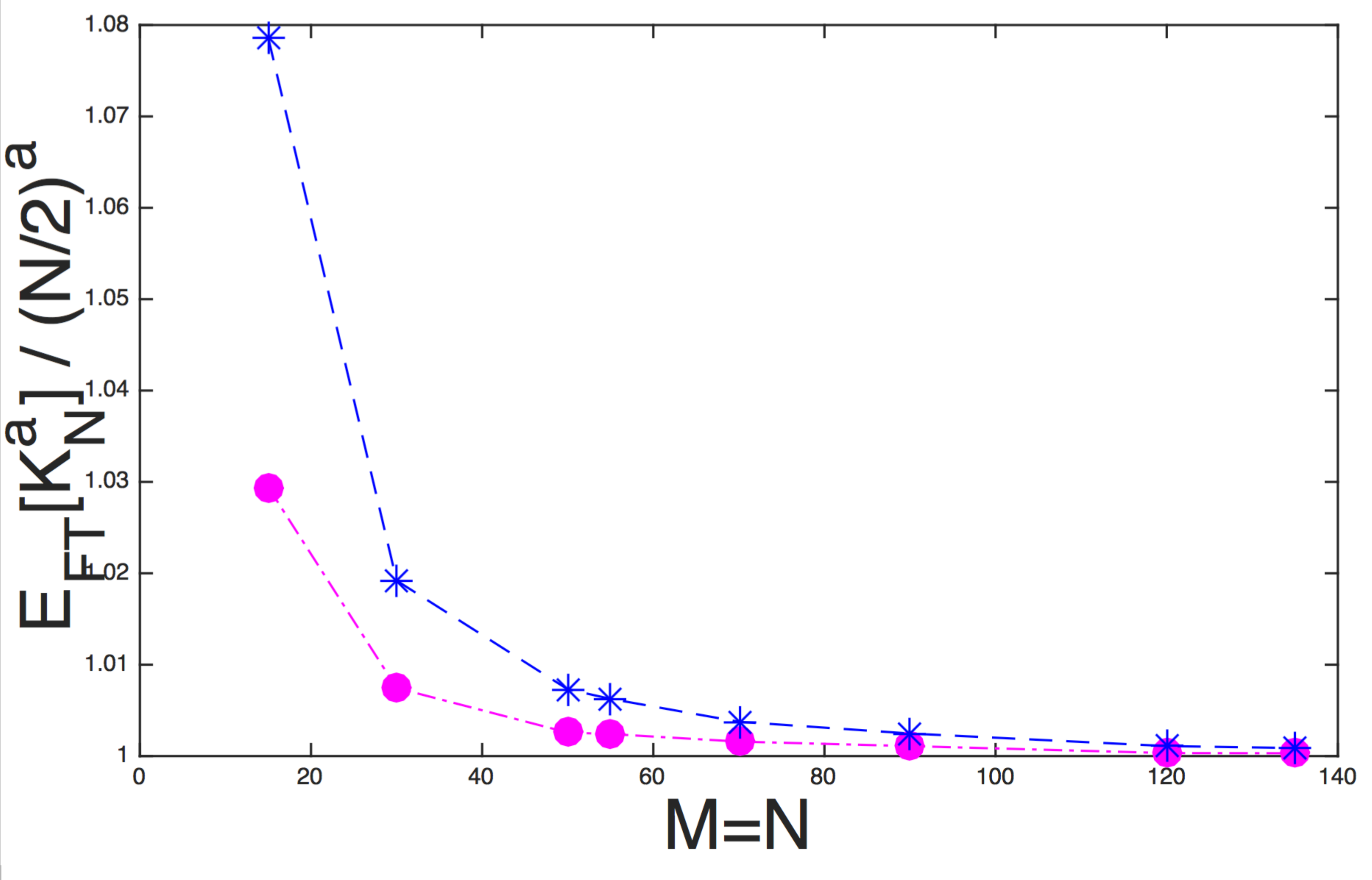}
\end{center}
\caption{(color online). $\mathbb{E}_{FT}\{K_N^a\}/(N/2)^a$ as a function of $M=N$. Numerical simulations for $a=3.2$ (purple dots) and $a=6.2$ (blue stars) confirm that the ratio converges to $1$, as predicted theoretically. }
\label{figsquare}
\end{figure}

In Appendix \ref{appendixD} we construct an exact asymptotic expansion of the integral in \eqref{approximationK}. We
realize, indeed, that the approximation based on \eqref{approximationK} allows us to deduce only the exact leading-$N$ behavior of $\mathbb{E}_{FT}\left\{ K^{a}\right\}$, while already the first subleading term is incorrect. The leading behavior,
 for $a \ll  N^2/2$ and $N \to \infty$, has the following form
\begin{equation}
\label{bigNK1}
\mathbb{E}_{FT}\left\{ K^{a}\right\} \sim \left(\dfrac{N}{2}\right)^{a}\ .
\end{equation}
This result, on the one hand, is somewhat trivial because it corresponds to the ultimate regime where the pdf of $K$ attains the form of the delta function. On the other hand, it nonetheless provides an important, physically meaningful information: namely, it states that for $N \to \infty$ the Schmidt number $K \sim N/2$ and hence, in contrast to the systems with fixed $N$ and $M \to \infty$, the entanglement in square systems is far from being complete. In Fig. \ref{figsquare}, we provide results of numerical diagonalization for the ratio $\mathbb{E}_{FT}\{K_N^a\}/(N/2)^a$ as a function of $M=N$, which show a convincing convergence to $1$ as predicted theoretically.

Finally, taking advantage once again of the general results due to Scott and Caves \cite{scott} and Giraud \cite{Giraud2}, for square $N = M$ systems
we present the large-$N$ asymptotics of the moments of the reciprocal of the Schmidt number - the purity $\Sigma_2$. We have that the first four moments obey
\begin{align}
\mathbb{E}\left\{\Sigma_2\right\} = \left(\dfrac{2}{N}\right) \left(1 - \dfrac{1}{N^2} + \dfrac{1}{N^4} + \mathcal{O}\left(\dfrac{1}{N^6}\right)\right)\ ,
\end{align}
\begin{align}
\mathbb{E}\left\{\Sigma_2^2\right\} = \left(\dfrac{2}{N}\right)^2 \left(1 - \dfrac{3}{2 \, N^2} - \dfrac{3}{2 \, N^4} + \mathcal{O}\left(\dfrac{1}{N^6}\right)\right)\ ,
\end{align}
\begin{align}
\mathbb{E}\left\{\Sigma_2^3\right\} = \left(\dfrac{2}{N}\right)^3 \left(1 - \dfrac{3}{2 \, N^2} - \dfrac{7}{ N^4} + \mathcal{O}\left(\dfrac{1}{N^6}\right)\right)\ ,
\end{align}
and
\begin{align}
\mathbb{E}\left\{\Sigma_2^4\right\} = \left(\dfrac{2}{N}\right)^4 \left(1 - \dfrac{1}{N^2} - \dfrac{57}{4 \, N^4} + \mathcal{O}\left(\dfrac{1}{N^6}\right)\right)\ ,
\end{align}
From \eqref{varsigmapurity} \cite{scott,Giraud1,Giraud2}, we have that 
the variance of the purity for square systems follows in the asymptotic limit $N \to \infty$
\begin{align}
{\rm Var}\left(\Sigma_2\right) =  \dfrac{2}{N^4} \left(1 - \dfrac{9}{N^2} + \dfrac{47}{N^4} + \mathcal{O}\left(\dfrac{1}{N^6}\right)\right) \ .
\end{align}
This means that the variance of the purity in square systems decays, to leading order, as an inverse fourth power of $N$, i.e. much faster than the variance of the von Neumann entropy (see our \eqref{squareentropyvar}). This implies, in turn, that the distribution of the purity is much more narrow than that of the von Neumann entropy.

\section{Conclusions}
\label{conc}

To summarize, we have analyzed the nonlinear statistics of Schmidt eigenvalues for entangled random pure states at finite $N,M$ - the dimensions of the Hilbert space partitions.
We have established 
a general relation between 
the $n$-point densities and the cross-moments of
 the $\beta$-FT ensemble and the analogous quantities of the $\beta$-WL ensembles. Building on these results, 
we presented explicit, closed-form expressions for the two-point densities and also an exact expression for the variance of the von Neumann entropy, valid for any $N$ and $M$.  

Further on, we derived a wealth of results 
for the Schmidt number $K$. 
Capitalizing on the known results for the distribution function 
of the purity \cite{Giraud1,Giraud2,majlarge,majlarge2},  
we calculated the probability density function of $K$ for $N = 2$ and $N=3$ and arbitrary $M$, and also discussed the forms of the right and left  tails of this distribution for square $N = M$ systems in the limit $N \to \infty$. 
From these results, we derived exact expression for the moments $\mathbb{E}\{K^a\}$ of $K$ of arbitrary order for $N = 2$ and $N = 3$ and arbitrary $M$, and analyzed their asymptotic large-$M$ behavior. 

Next, taking advantage 
of the established relation between the cross-moments of the FT and WL ensembles,  we found an exact representation of $\mathbb{E}\{K^a\}$ of arbitrary, not necessarily integer order $a$ in $N \times N$ systems by spotting a previously unnoticed connection with the statistics of the smallest eigenvalue of Gaussian Unitary matrices. Lastly, we discussed the asymptotic, large-$N$ behavior of these moments. Our results have been corroborated via numerical simulations whenever possible, with excellent agreement.     

As a by-product of our analysis, we also established an exact asymptotic expansion 
of the probability $P(x_{min}^{GUE} \geq \sqrt{2 N} \xi)$ that the smallest eigenvalue in the Gaussian Unitary Ensemble 
is larger than $\sqrt{2 N} \xi$, in the limit $\xi \to \infty$ for fixed $N$, by identifying the coefficients in this expansion via the moments of the purity in the fixed-trace ensemble.

\section*{Acknowledgements}

The authors acknowledge helpful discussions with D.S. Dean, O. Giraud and S. N. Majumdar. 
PV acknowledges
the stimulating research environment provided by the EPSRC Centre for Doctoral
Training in Cross-Disciplinary Approaches to Non-Equilibrium Systems (CANES,
EP/L015854/1).
GO acknowledges a partial support from the ONRG Grant N62909-15-1-C076 and wishes to
 thank for warm hospitality and financial support the Institute for Mathematical Sciences of 
 the National University of Singapore, where some part of this work has been done.

\vspace{2in}

\appendix

\section{\label{appendix} Coefficients $A^{(1)}_l(N,M)$ and $A^{(2)}_{l,j}(N,M)$.}

The spectral density $\rho_1^{(WL)}(y)$ of the $\beta=2$ WL ensemble reads (see, e.g., \cite{livan})
\begin{align}
\label{LL}
&\rho_1^{(WL)}(y) = \frac{y^{b}}{2^{b+1} N} \,  e^{-y/2} \sum_{m=0}^{N - 1} \frac{m!  \left(L_m^{(b)}(\frac{y}{2})\right)^2}{(m+b)!} \nonumber\\
&=  \frac{\Gamma(N) \, y^{b}}{2^{b+1} \Gamma(M)} e^{-y/2} 
\Big[ L_{N-1}^{(b)}\left(\frac{y}{2}\right) \, L_{N-1}^{(b+1)}\left(\frac{y}{2}\right) \nonumber\\
&-  L_{N}^{(b)}\left(\frac{y}{2}\right) \, L_{N-2}^{(b+1)}\left(\frac{y}{2}\right)\Big] \,, \nonumber\\
\end{align}
where $L_m^{(b)}(\cdot)$ is the generalized Laguerre polynomial. Further on, we use the following representation of the product of two generalized Laguerre polynomials 
\begin{align}
\label{L}
L_m^{(a)}(x) L_n^{(b)}(x) &= \sum_{l=0}^{m+n} (-1)^l \frac{x^l}{l!} \sum_{p=0}^m \binom{l}{p} \nonumber\\ 
&\times \binom{m+a}{m-p} \binom{n+b}{n-l+p} \ ,
\end{align}
where $\binom{l}{p}$ is the binomial coefficient such that $\binom{l}{p} = l!/p! (l-p)!$ for $l \geq p$ and $p \geq 0$, and zero otherwise. Inserting \eqref{L} into \eqref{LL}, and collecting terms with the same power of $y$,
we find the following explicit representation of the coefficients   $A_l^{(1)}(N,M)$ in \eqref{j} 
\begin{align}
\label{a1}
A_l^{(1)}(N,M) &= \frac{(-1)^l}{2^{b+l+1} l!} \frac{\Gamma(N)}{\Gamma(M)}  \sum_{p=0}^{N} \binom{l}{p} \nonumber\\
&\times\left[ \binom{M-1}{b+p} \binom{M}{b+1+l-p} -\right. \nonumber\\
&\left.-\binom{M}{b+p} \binom{M-1}{b+1+l-p}\right] \,.
\end{align}

Further on, we present the derivation of the coefficients $A^{(2)}_{l,j}(N,M)$.
The normalized two-point density of the $\beta=2$ WL ensemble is given explicitly by
\begin{align}
\label{n}
&\rho_2^{(WL)}(y_1,y_2) = \frac{N}{N-1} \Big( \rho_1^{(WL)}(y_1) \rho_1^{(WL)}(y_2) - \nonumber\\
&-\frac{(y_1 y_2)^{b}}{4^{b+1} N^2} e^{-y_1/2 - y_2/2 } \times \nonumber\\
&\times \Big(\sum_{m=0}^{N - 1} \frac{m!}{(m + b)!} L_{m}^{(b)}\left(\frac{y_1}{2}\right) L_{m}^{(b)}\left(\frac{y_2}{2}\right)\Big)^2\Big) \nonumber\ .\\
\end{align}
The first term in square brackets, which is a product of two one-point densities, 
produces a trivial contribution to $A_{l,j}^{(2)}(N,M)$ - the product of two corresponding 
coefficients $A_l^{(1)}(N,M)$. We therefore focus on the second term. The sum of products of Laguerre polynomials entering \eqref{n} can be written down as
\begin{align}
\label{t}
&\sum_{m=0}^{N-1} \frac{m!}{(m+b)!} L_m^{(b)}\left(\frac{y_1}{2}\right) L_m^{(b)}\left(\frac{y_2}{2}\right) = \nonumber\\
&= \sum_{f,r=0}^{N-1} C_{f,r}(N,M) \, y_1^r \, y_2^f  \,,
\end{align}
where the numerical coefficients $C_{f,r}(N,M)$ are given by
\begin{equation}
C_{f,r}(N,M) = \frac{(-1)^{f+r}}{2^{f+r} f! (b+r)!} \sum_{m=0}^{N-1} \binom{m}{r} \binom{m+b}{m-f} \,.
\end{equation}
Consequently, the squared sum on the left-hand-side of \eqref{t} admits the following expansion
\begin{align}
&\left(\sum_{m=0}^{N-1} \frac{m!}{(m+b)!} L_m^{(b)}\left(\frac{y_1}{2}\right) L_m^{(b)}\left(\frac{y_2}{2}\right)\right)^2 = \nonumber\\
&= \sum_{l,j=0}^{2 N - 2} K_{l,j}(N,M) \, y_1^l \, y_2^j\ ,
\end{align}
where
\begin{equation}
K_{l,j}(N,M) = \sum_{r_1,f_1=0}^{N - 1} C_{f_1,r_1}(N,M) \, C_{j-f_1,l-r_1}(N,M) \,.
\end{equation}
Consequently, we arrive at the following explicit result for the coefficients $A_{l,j}^{(2)}(N,M)$
\begin{align}
\label{a2}
A_{l,j}^{(2)}(N,M) &= \frac{N}{N-1} A_{l}^{(1)}(N,M) A_{j}^{(1)}(N,M) - \nonumber\\
&- \frac{K_{l,j}(N,M)}{4^{b+1} N (N-1)} \,.
\end{align}

\section{\label{appendix3} Moments of $K$ for $N = 3$}

For $N = 3$ and arbitrary $M \geq 3$, it is convenient to calculate the moments of $K$ directly from their formal definition:
\begin{align}
\label{formal}
&\mathbb{E}_{FT}\left\{ K_{N=3}^{a}\right\} = \dfrac{\Gamma\left(3 M\right)}{12 \Gamma\left(M\right) \Gamma\left(M - 1\right) \Gamma\left(M-2\right)} \nonumber\\ 
&\times \int_0^1 \int^1_0 \int^1_0 d\ld_1 d\ld_2 d\ld_3 \, \left(\ld_1 \ld_2 \ld_3\right)^{M-3}
 \nonumber\\
&\times \dfrac{ |\triangle({\bm \ld})|^{2}}{\left(\ld_1^2 + \ld_2^2 + \ld_3^2\right)^{a}} \, \delta\left(\ld_1+\ld_2+\ld_3 - 1\right)\ .
\end{align}
One of the integrals, say, over $\ld_3$, can be simply performed using the delta function. Changing then the integration variable $\ld_2$ as
\begin{equation}
\ld_2 = \dfrac{(1 -\ld_1)}{2}\left(1 - \sqrt{1 - x}\right) \,,
\end{equation}  
we cast the expression in \eqref{formal} into the form
\begin{align}
\label{formal2}
&\mathbb{E}_{FT}\left\{ K_{N=3}^{a}\right\} = \dfrac{\Gamma\left(3 M\right)}{6 \, 2^{2 M} \Gamma\left(M\right) \Gamma\left(M - 1\right) \Gamma\left(M-2\right)} \nonumber\\
&\times \int^{1}_0 d \ld_1 \, \ld_1^{M - 3} \, \left(1-\ld_1\right)^{2 M - 3} 
 \int^1_0 dx \, x^{M-3} \, \left(1-x\right)^{1/2} \,   \nonumber\\
&\times \dfrac{ \left(4 \ld_1^2 - 4 \ld_1 \left(1-\ld_1\right) +\left(1- \ld_1\right)^2 x\right)^2}{\left(\ld_1^2 + \left(1 - \ld_1\right)^2 \left(1 - \dfrac{x}{2}\right)\right)^{a}} \,.
\end{align}
The integrals in \eqref{formal2} are coupled via the expression in the 
denominator of the kernel and in order to factorise them we use the following 
 expansion
\begin{align}
&\dfrac{1}{\left(\ld_1^2 + \left(1 - \ld_1\right)^2 \left(1 - \dfrac{x}{2}\right)\right)^{a}} = \sum_{n=0}^{\infty} \binom{n+a-1}{n} 2^n \nonumber\\
& \times \, \sum_{k=0}^n \binom{n}{k} (-1)^k (1- \ld_1)^{n+k} \left(1 - \dfrac{x}{4}\right)^k \,.
\end{align}
Plugging the latter expansion in \eqref{formal2} and performing the integration, we obtain the following result:
\begin{align}
\label{formal3}
&\mathbb{E}_{FT}\left\{ K_{N=3}^{a}\right\} = \dfrac{\sqrt{\pi} \, \Gamma\left(3 M\right)}{3 \, 2^{2 M} \Gamma\left(M-1\right)}
\sum_{n=0}^{\infty} \binom{n+a-1}{n} 2^n \nonumber\\
&\sum_{k=0}^n \binom{n}{k} (-1)^k \left(R^{(1)}_{n,k} + R^{(2)}_{n,k}  + R^{(3)}_{n,k} \right) \,,
\end{align}
where
\begin{align}
\label{r1}
&R^{(1)}_{n,k} = 4 \dfrac{\left(k^2 + k \left(2 M + 2 n -3\right) +\left(M+n\right)^2 - M - 3 n  + 2\right)}{\Gamma\left(M-\dfrac{1}{2}\right) \, \Gamma\left(3 M + n + k\right)}\nonumber\\
& \times \Gamma\left(2 M + k + n - 2\right)  \,_2F_1\left(-k, M-2; M-\dfrac{1}{2}; \dfrac{1}{4}\right) \,,
\end{align}
\begin{flalign}
\label{r2}
&R^{(2)}_{n,k} = \dfrac{\Gamma\left(2 M + n + k + 2\right)}{4 \, \Gamma\left(M + \dfrac{3}{2}\right) \, \Gamma\left(3 M + k + n\right)} \nonumber\\
& \times \,_2F_1\left(-k, M; M+\dfrac{3}{2}; \dfrac{1}{4}\right) \,,
\end{flalign}
and
\begin{align}
\label{r3}
&R^{(3)}_{n,k} = - 2 \dfrac{\left(M-2\right) \, \left(M+n+k+1\right)}{\left(M-1\right) \, \Gamma\left(M+\dfrac{1}{2}\right)} \dfrac{\Gamma\left(2 M + n + k\right)}{\Gamma\left(3 M + k + n\right)} \nonumber\\
& \times \,_2F_1\left(-k, M-1; M+\dfrac{1}{2}; \dfrac{1}{4}\right) \,.
\end{align}
Lastly, we note that the hypergeometric functions entering \eqref{r1} to \eqref{r3} can be simply expressed via the Jacobi polynomials $P_k^{(\alpha,\beta)}(x)$ with the argument $x=1/2$:`
\begin{align}
&\,_2F_1\left(-k,M-p;M-p+\dfrac{3}{2};\dfrac{1}{4}\right) = \dfrac{k! \, \Gamma\left(M-p+\dfrac{3}{2}\right)}{\Gamma\left(M-p+\dfrac{3}{2}+k\right)} \nonumber\\
&\times P_k^{(M-p+1/2,-k-3/2)}\left(\dfrac{1}{2}\right) 
=  \dfrac{k! \, \Gamma\left(M-p+\dfrac{3}{2}\right)}{\Gamma\left(M-p+\dfrac{3}{2}+k\right)} \nonumber\\
&\times \dfrac{(-1)^k}{4^k} \sum_{m=0}^k \binom{k+M-p+1/2}{m} \binom{-3/2}{k-m} (-3)^m \,,
\end{align}
where $p=0,1,2$
Therefore, our \eqref{formal3} with \eqref{r1} to \eqref{r3} defines an exact result for the moments  
$\mathbb{E}_{FT}\left\{ K_{N=3}^{a}\right\}$ of the Schmidt number for $N=3$ and arbitrary $M$ and (not necessarily integer) $a$ in form of an infinite series. In principle, summation over $k$ and $n$ can be performed giving an explicit result in terms of a (rather cumbersome) combination of generalized hypergeometric functions. 
For any fixed $a$ and $M$, this series can be straightforwardly computed using Mathematica. 

The expression in \eqref{formal3} is, however, not very useful since it has a too complicated structure and does not permit to easily observe the $M$- and $a$-dependence of the moments of $K$. To this purpose, we focus next on the asymptotic, large-$M$ behavior of the expression in \eqref{formal3}.  Expanding the ratios of the gamma functions entering \eqref{r1} to \eqref{r3} in Taylor series in inverse powers of $M$, and taking advantage of the following asymptotic expansion of the
hypergeometric functions for large values of the parameters,
 \begin{align}
 \label{asympF}
& \,_2F_1\left(-k,M+p-2; M+p-2+3/2;z\right) = \nonumber\\
&= \left(1-z\right)^{k} \left[1 + \dfrac{3 k z}{2 (1 - z) M} + \nonumber\right.\\
 &\left.+ \dfrac{3 k z \left(2 - 4 p (1-z) - 7 z + 5 k z\right)}{8 (1 - z)^2 M^2} + \mathcal{O}\left(\dfrac{1}{M^3}\right)
  \right] \nonumber\ , \\
\end{align}
we have
\begin{align}
\label{asia}
&\dfrac{\sqrt{\pi} \, \Gamma\left(3 M\right)}{3 \, 2^{2 M} \Gamma\left(M-1\right)} \, \sum_{k=0}^n \binom{n}{k} (-1)^k \left(R^{(1)}_{n,k} + R^{(2)}_{n,k}  + R^{(3)}_{n,k} \right) \nonumber\\
&=\dfrac{1}{3^n} \left(1 - \dfrac{7 \, n}{6 \, M} + \dfrac{7 \, n \, (9 n-5)}{72 \, M^2} + \mathcal{O}\left(\dfrac{1}{M^3}\right) \right) \,.
\end{align}
Multiplying the last line in \eqref{asia} by $2^n \binom{n+a-1}{n}$ and summing over $n$, we arrive at the result in \eqref{momN=3}.

\section{\label{appendixC} An alternative derivation of the moments of the Schmidt number of order $a < N^2/2$.}

In this Appendix, we present an alternative derivation of the expression \eqref{expressionK} defining moments of the Schmidt number $K$ of order $a < N^2/2$. Our approach here is, in fact, 
essentially the same as the one we developed 
for the derivation of the $n$-point densities of the FT ensemble.

First of all, we introduce an auxiliary function 
\begin{align}
\label{purity}
&P(\Sigma_2,t) = Z^{-1}_{N,N} \int^{\infty}_0 \ldots \int^{\infty}_0 \delta\left(\sum_{i=1}^N \ld_i - t\right) \, \nonumber\\
&\times \delta\left(\Sigma_2 - \sum_{i=1}^N \ld_i^2\right) \,\triangle^2({\bm \ld}) \, \prod_{i=1}^N d\ld_i \ ,
\end{align}
(with $Z^{-1}_{N,N}$  defined in \eqref{norm} with $N = M$, $\beta=2$ and hence, $\mu = N^2$) which describes for $t=1$ the pdf of the purity $\Sigma_2$ on $N \times N$ FT ensembles with $\beta = 2$, see \eqref{jpdf11}. Taking the double Laplace transform 
of the expression in \eqref{purity} with respect to both $t$ and $\Sigma_2$, we have
\begin{align}
&{\cal F}(z,p) = {\cal L}_{p,t}\left({\cal L}_{z,\Sigma_2}\left(P(\Sigma_2,t)\right)\right) \nonumber\\
&= Z^{-1}_{N,N} \int^{\infty}_0 \ldots \int^{\infty}_0 \triangle^2({\bm \ld}) \, \nonumber\\
&\times \exp\left(- p \sum_{i=1}^N \ld_i - z \sum_{i=1}^N \ld_i^2\right) \, \prod_{i=1}^N d\ld_i
\end{align}
Changing the integration variables $x_i = \sqrt{z} \ld_i + p/2 \sqrt{z}$, we can cast the latter expression into the form
\begin{align}
{\cal F}(z,p) &=  Z^{-1}_{N,N} \exp\left(\frac{p^2 N}{4 z}\right) z^{-N^2/2} \int^{\infty}_{p/2 \sqrt{z}} \ldots \int^{\infty}_{p/2 \sqrt{z}} \nonumber\\
&\times  \triangle^2({\bm x}) \, \exp\left(- \sum_{i=1}^N x_i^2\right) \, \prod_{i=1}^N dx_i \ ,
\end{align}
which can be immediately 
written in the following more appealing form
\begin{align}
\label{f}
{\cal F}(z,p) &=  Z^{-1}_{N,N} Z_{N,N}(GUE) \, z^{-N^2/2}  \, \nonumber\\
&\times \exp\left(\frac{p^2 N}{4 z}\right)  \, P\left(x_{min}^{GUE} \geq \frac{p}{2 \sqrt{z}}\right) \ . \nonumber\\
\end{align}
Here, $ Z_{N,N}(GUE) $ is the inverse of the normalization constant of the $N \times N$ Gaussian Unitary Ensemble, defined explicitly in \eqref{normGUE}, and $P(x_{min}^{GUE} \geq p/2 \sqrt{z})$ is as before the probability that the smallest eigenvalue in this ensemble is greater or equal to $p/2 \sqrt{z}$. 
Consequently, the pdf of the purity can be formally written down as
\begin{align}
P(\Sigma_2) &= Z^{-1}_{N,N} Z_{N,N}(GUE)  \nonumber\\
&\times {\cal L}_{t=1,p}^{-1}\Big({\cal L}^{-1}_{\Sigma_2,z}\Big( z^{-N^2/2} \times  \nonumber\\
&\times  \exp\left(\frac{p^2 N}{4 z}\right) \, P\left(x_{min}^{GUE} \geq \frac{p}{2 \sqrt{z}}\right)\Big)\Big) \ .
\end{align}
Note now that the moments of the Schmidt number $K$ can be straightforwardly expressed in terms of the function ${\cal F}(z,p)$ as
\begin{align}
&\mathbb{E}_{FT}\left\{ K^{a}\right\} = \frac{1}{\Gamma(a)} {\cal L}_{t=1,p}^{-1}\left(\int^{\infty}_0 dz\ z^{a-1}  {\cal F}(z,p)\right) \nonumber\\
&= \frac{Z^{-1}_{N,N} Z_{N,N}(GUE) }{\Gamma(a)}   {\cal L}_{t=1,p}^{-1}\Big(\int^{\infty}_0 z^{a-1-N^2/2} \nonumber\\
&\times  \exp\left(\frac{p^2 N}{4 z}\right) P\left(x_{min}^{GUE} \geq \frac{p}{2 \sqrt{z}}\right)\ dz\Big)\ .
\end{align}
Changing the integration variable $\xi = p/2 \sqrt{2 N z}$, we formally rewrite the latter expression as
\begin{align}
&\mathbb{E}_{FT}\left\{ K^{a}\right\} =   \frac{2 Z^{-1}_{N,N} Z_{N,N}(GUE) (8 N)^{N^2/2-a}}{\Gamma(a)}  \nonumber\\
&\times {\cal L}_{t=1,p}^{-1}\left(\frac{1}{p^{N^2 - 2 a}}\right) \times \nonumber\\
&\times \int^{\infty}_0 d\xi \, \xi^{N^2 - 2 a - 1} e^{2 N^2 \xi^2} \, P(x_{min}^{GUE} \geq \sqrt{2N} \xi) \ ,
\end{align}
which gives upon the inversion of the Laplace transform with respect to $p$, (for $a < \mu/2$), our result in \eqref{expressionK}.

\section{\label{appendixD} Asymptotic large-$N$ behavior of the moments of the Schmidt number for square systems}

Here we detail the evaluation  of the integral 
in  \eqref{approximationK}, which defines the moments of $K$ of arbitrary order $a < N^2/2$ under the assumption that the probability
$P(x_{min}^{GUE} \geq \sqrt{2 N} \xi)$ can be well approximated 
by its large deviation form calculated in \cite{DM}.  Our aim is to present the first three terms in the asymptotic large-$N$ expansion stemming out of this approximation. 

We start with an analysis of the behavior of the integrand in
 \eqref{approximationK}. The integrand, i.e.,  
\begin{equation}
f(\xi) = \xi^{N^2 - 2 a - 1} \exp\left(2 N^2 \left(\xi^2 - \Phi(\xi)\right)\right)
\end{equation}
is a bell-shaped function of $\xi$, which vanishes for $\xi=0$ and $\xi \to \infty$, and has a maximum at
\begin{align}
\xi_{max} = \left(\frac{N^3 + \left(N^2 + 6 a + 3\right)^{3/2} - 9 (1 + 2 a) N}{4 (1+ 2 a) N}\right)^{1/2}\ . \nonumber\\
\end{align}
The position of the maximum $\xi_{max}$ is a monotonically increasing function of $N$, for a fixed $a$, and is a monotonically 
decreasing function of $a$, at a fixed $N$, for $a \in [0,N^2/2]$;  
$\xi_{max}(a=N^2/2) \approx 3 \sqrt{3}/8 N^2 \to 0$ for $N \to \infty$. 

Given that there is a large parameter $N^2$ in the exponential, it is tempting to resort to the saddle-point approximation. This approach, however, yields a very poor result for the integral. The reason is that, even though $f(\xi)$ can be very well approximated by a Gaussian in the vicinity of $\xi=\xi_{max}$, this Gaussian is flanked on both sides by power-law tails : 
\begin{equation}
\label{righttail}
f(\xi) \approx \xi^{N^2 - 2 a - 1}
\end{equation} 
when $\xi \to 0$,
since
\begin{equation}
\xi^2 - \Phi(\xi) = - \frac{\ln 3}{4} - \frac{4 \xi}{ 3 \sqrt{3}} + \mathcal{O}\left(\xi^2\right) \,
\end{equation}
in this limit, and using the expansion in \eqref{largexi}, 
\begin{equation}
\label{rightful}
f(\xi) \approx \frac{1}{\xi^{2 a + 1}} \,,
\end{equation}
as $\xi \to \infty$.  As a matter of fact, these power-law tails 
provide the dominant contributions to the integral in \eqref{approximationK}; namely, 
 the left tail dominates the behavior of the moments of $K$ when $a$ is close to $N^2/2$, and the right tail does the same for sufficiently small values of  $a$. 
Hence, we have to resort
here to a different approach. 

Lastly, 
we note that there is some subtlety in the behavior of the integral in \eqref{approximationK} (and also in the exact expression in \eqref{expressionK}) in the limit $a \to 0$, since in this case $f(\xi) \approx 1/\xi$, (see \eqref{rightful}). This implies that
the integrals are formally logarithmically divergent on the upper limit of integration. On the other hand,  the expressions in \eqref{expressionK} and \eqref{approximationK} contain a factor $1/\Gamma(a)$, which vanishes as $a \to 0$. As a matter of fact, these two conflicting factors compensate each other and $\lim_{a \to 0} \mathbb{E}_{FT}\{K^{a}\}$ exists, when one first performs the integrals at a fixed $a >0$ and only afterward takes the limit $a \to 0$.

Now, we turn to the evaluation of the integral in \eqref{approximationK}. Changing the integration variable as
\begin{equation}
\label{change}
\xi = \dfrac{\sqrt{3}}{2} \dfrac{(1-z)}{\sqrt{z}} \,
\end{equation}
the expression for the moments of $K$ 
in \eqref{approximationK} can be conveniently
cast into the form 
\begin{align}
\label{approximationK4}
&\mathbb{E}_{FT}\left\{ K^{a}\right\} \sim \frac{(2 \pi)^{N/2} N^{N^2/2}}{(6 N)^{a} G(N+1)} \frac{\Gamma(N^2)}{\Gamma(a) \Gamma(N^2- 2 a)} C_N  \nonumber\\
&\times\int^1_0 dz (1+z) z^{a-1} \left(1-z\right)^{N^2 - 2 a - 1} \nonumber\\
&\times \exp\left(- \dfrac{3 N^2}{4} (1-z)\left(1+\dfrac{z}{9}\right)\right) \,.
\end{align}
Further on, using the well-known expression for the generating function of the Hermite polynomials $H_m(\ldots)$,
we represent the exponential in the latter equation as a sum over Hermite polynomials,
\begin{align}
&\exp\left(- \dfrac{3 N^2}{4} (1-z)\left(1+\dfrac{z}{9}\right)\right) = \exp\left(- \dfrac{3 N^2}{4}\right) \nonumber\\
&\times \sum_{m=0}^{\infty} \dfrac{(-1)^m}{m!} H_m\left(\dfrac{2 i N}{\sqrt{3}}\right) \left(\dfrac{i N z}{2 \sqrt{3}}\right)^m \,,
\end{align}
where $i = \sqrt{-1}$.
Inserting this expression into \eqref{approximationK4} and performing the integral, we arrive at the following representation of the moments of $K$ in the form of an infinite series
\begin{align}
\label{expressionK3}
&\mathbb{E}_{FT}\left\{ K^{a}\right\} \sim \frac{(2 \pi)^{N/2} N^{N^2/2} \Gamma(N^2)}{(6 N)^{a}  \Gamma(a) G(N+1)} \exp\left(- \dfrac{3}{4} N^2\right)  \nonumber\\
&\times C_N \sum_{m=0}^{\infty} \dfrac{(-1)^m (N^2+2 m)}{m!} \dfrac{\Gamma\left(a+m\right)}{\Gamma\left(N^2-a+m+1\right)} \nonumber\\
&\times H_m\left(\dfrac{2 i N}{\sqrt{3}}\right) \left(\dfrac{i N}{2 \sqrt{3}}\right)^m \,,
\end{align} 
which is completely equivalent to the expression in \eqref{approximationK}. 

\begin{figure}[hb]
\centerline{\includegraphics[width=8cm]{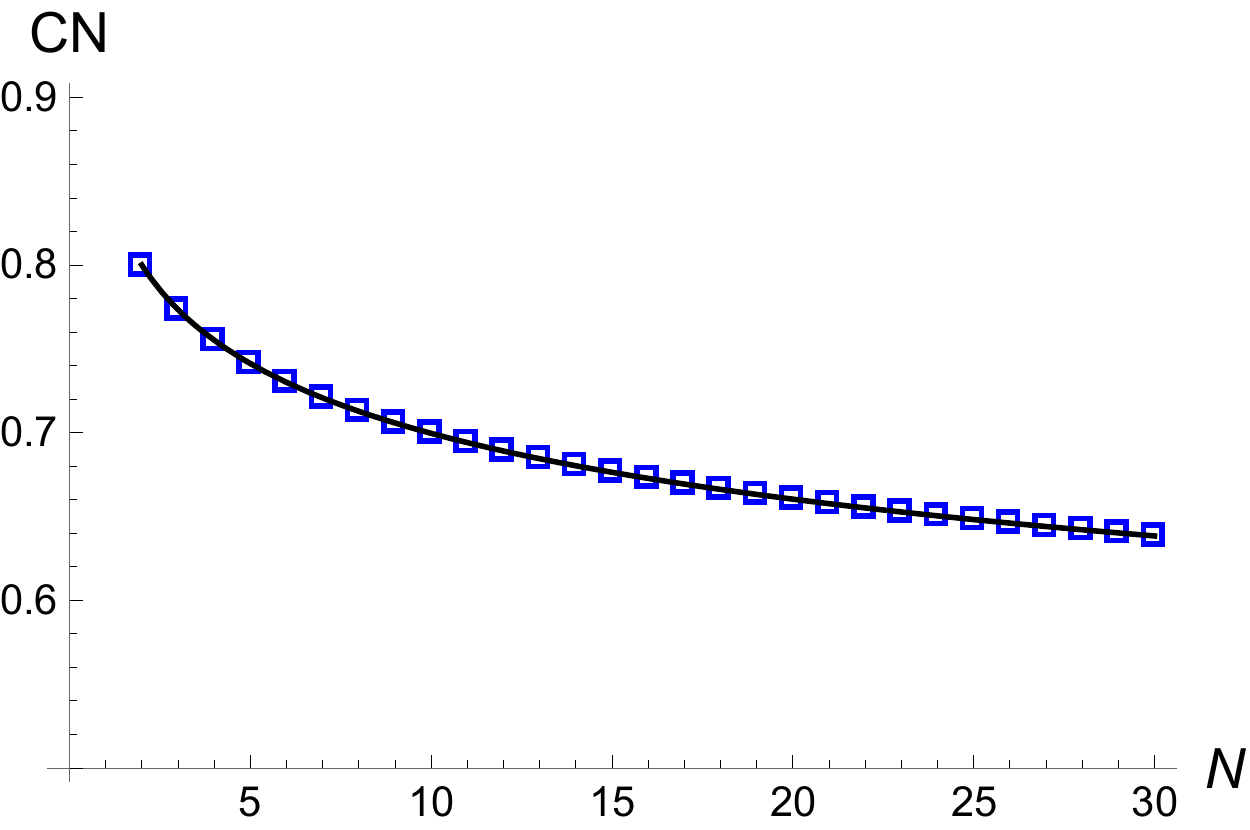}}
\caption{(color online). Normalization constant $C_N$, \eqref{CNn} (symbols), as a function of $N$. The solid curve is the asymptotic result in \eqref{CNasymp}. }
\label{CNN}
\end{figure}

We are now in the position to determine the $N$-dependent constant $C_N$. 
Taking in \eqref{expressionK3} the limit $a \to 0$, and requiring that $\lim_{a \to 0} \mathbb{E}_{FT}\left\{ K^{a}\right\} \equiv 1$, we find the expression in \eqref{CNn}.
The normalization constant $C_N$ as a function of $N$ is depicted in Fig. \ref{CNN}. One observes that, indeed, $C_N$ is a very slowly varying function of $N$. As a matter of fact, the asymptotic behavior of $C_N$ is well described by a slow power law of the form
\begin{equation}
\label{CNasymp}
C_N \approx \dfrac{1}{A} \left(\dfrac{e}{N}\right)^{1/12} \,,
\end{equation}
where $A \approx 1.282$ is the Glaisher's constant. One infers from Fig. \ref{CNN} that this asymptotic form
sets in starting from very moderate values of $N$. Moreover, we note that $C_N$ defines precisely the terms in the second line in \eqref{ququ}.

Next, we take advantage of the explicit representation of the Hermite polynomials
\begin{equation}
\label{hermit}
H_m\left(x\right) = m! \sum_{k=0}^{\lfloor m/2 \rfloor} \dfrac{(-1)^k}{k! (m- 2k)!} \left(2 x\right)^{m-2 k} \,,
\end{equation}
where $\lfloor ... \rfloor$ is the floor function. Note that the series in \eqref{hermit} are arranged in descending order with respect to powers of the argument, so that the term $k=0$ corresponds to the highest power of $x$. Inserting \eqref{hermit} into 
\eqref{expressionK3}, taking into account \eqref{CNn} and performing summation over $m$, we have
\begin{align}
\label{expressionK7}
&\mathbb{E}_{FT}\left\{ K^{a}\right\} \sim \dfrac{\Gamma\left(N^2\right)}{(6 N)^{a} \Gamma(a)} \sum_{k=0}^{\infty} \dfrac{\left(N/2\sqrt{3}\right)^{2 k}}{k!}  \nonumber\\
&\times \Big[ \dfrac{\left(N^2 + 4 k\right) \Gamma\left(a + 2 k\right)}{\Gamma\left(N^2 - a + 2 k + 1\right)} \nonumber\\
&\times \, _1F_1\left(a+ 2k, N^2 - a + 2 k +1, \dfrac{2 N^2}{3}\right) + \nonumber\\
&+ \dfrac{4 N^2}{3} \dfrac{\Gamma(a + 2 k + 1)}{\Gamma\left(N^2 - a + 2 k + 2\right)} \nonumber\\
&\times \, _1F_1\left(a+ 2k+1, N^2 - a + 2 k +2, \dfrac{2 N^2}{3}\right)
\Big] \,,
\end{align}
where $_1F_1(\ldots)$ is the Kummer's confluent hypergeometric function.  
 
Note that in a similar fashion we can obtain an explicit expression for the average logarithm of $K$, which describes the "typical" behavior of the Schmidt number. Using our \eqref{expressionK3}, we find
\begin{align}
\label{avlog}
&\mathbb{E}_{FT}\left\{ \ln(K)\right\} = \lim_{a \to 0} \dfrac{1}{a} \left(E_{FT}\left\{ K^{a}\right\} - 1\right)  \nonumber\\
&\sim \psi^{(0)}\left(N^2+1\right) -  \ln(6 N) + \nonumber\\
 &+ \Gamma\left(N^2\right) \sum_{m=1}^{\infty} \dfrac{(-1)^m}{m} \dfrac{\left(N^2 + 2 m\right)}{\Gamma\left(N^2 + m + 1\right)} \nonumber\\
 &\times H_m\left(\dfrac{2 i N}{\sqrt{3}}\right) \left(\dfrac{i N}{2 \sqrt{3}}\right)^m \,,
\end{align}
where $\psi^{(0)}(\ldots)$
is the digamma function, defined in the text after \eqref{mean2}.

We turn to the analysis of the asymptotic behavior of the result in \eqref{expressionK7}   in the limit $N \to \infty$.
We note first that the Kummer's functions entering the series have the form $_1F_1(b,c,x)$, in which $b$ is independent of $N$, while the parameter $c$ and the argument $x$ are both proportional to $N^2$, and hence, tend to infinity as $N \to \infty$. Setting $x = \zeta c$, where in the case at hand $\zeta$ is a bounded function such that $\zeta < 1$, we have that in the limit $c \to \infty$ the Kummer's functions obey
\begin{align}
\label{kummer}
&_1F_1(b,c, \zeta c) = \dfrac{1}{\left(1-\zeta\right)^b} \Big[1 - \dfrac{b (b+1) \zeta^2}{2 (1 - \zeta)^2 c}\Big(1 - \nonumber\\
&- \dfrac{\left(12 + 8 (1+2 b)\zeta +(b-1)(3 b+2) \zeta^2\right)}{12 (1 - \zeta)^2 c}\Big) + \mathcal{O}\left(\dfrac{1}{c^3}\right)\Big] \,,\nonumber\\
\end{align}
which implies that all $_1F_1(b,c,x)$ in \eqref{expressionK7} approach constant values as $N \to \infty$, and the dominant $N$-dependence of each term in the expansion in \eqref{expressionK7} will come from the ratio of the gamma functions. 
Noticing next that as $N \to \infty$,
\begin{equation}
\dfrac{\Gamma\left(N^2\right) N^{2 k + 2}}{N^{a} \Gamma\left(N^2 - a + 2k + 1\right)} = \mathcal{O}\left(N^{a - 2k}\right) \,,
\end{equation}
and 
\begin{equation}
\dfrac{\Gamma\left(N^2\right) N^{2 k + 2}}{N^{a} \Gamma\left(N^2 - a + 2k + 2\right)} = \mathcal{O}\left(N^{a - 2k - 2}\right) \,,
\end{equation}
we can conclude that for large $N$ the series in \eqref{expressionK7} represents an expansion in the inverse powers of $N^2$. Moreover, the dominant contribution to the large-$N$ behavior of  $\mathbb{E}_{FT}\left\{ K^{a}\right\}$ will be provided by the zeroth term, while the terms of higher order will contribute only to the subdominant behavior.  More precisely, we have that the zeroth term ($k=0$) of the series is explicitly given by
\begin{align}
\label{cont0}
&\left(\dfrac{N}{2}\right)^{a} \Big[1 + \dfrac{a (1-a)}{2 N^2} + \dfrac{a (158 + 249 a + 70 a^2 + 3 a^3)}{24 N^4} + \nonumber\\ &+ \mathcal{O}\left(\dfrac{1}{N^6}\right)\Big] \,,
\end{align}
the first one ($k=1$) obeys
\begin{align}
\label{cont1}
&\left(\dfrac{N}{2}\right)^{a} \Big[\dfrac{3 a (1+a)}{4 N^2} - \dfrac{3 a(1+a) (5+a)(6+a)}{8 N^4} + \nonumber\\
&+ \mathcal{O}\left(\dfrac{1}{N^6}\right)\Big] \,,
\end{align}
while the second one ($k=2$) (contributing only to the order $\mathcal{O}(1/N^4)$) is explicitly given by
\begin{align}
\label{cont2}
\left(\dfrac{N}{2}\right)^{a} \Big[\dfrac{9 a (1+a) (2+a)(3+a)}{32 N^4} + \mathcal{O}\left(\dfrac{1}{N^6}\right)\Big] \,.
\end{align}
The terms with higher values of $k$ contribute to the order $\mathcal{O}(1/N^6)$ and higher, and hence can be safely neglected, given that we are interested in the large-$N$ behavior of the moments of $K$ to the order $\mathcal{O}(N^{a-4})$ at most.  Summing the contributions given in \eqref{cont0}, \eqref{cont1} and \eqref{cont2}, we arrive at the following result
\begin{align}
\label{bigNK}
&\mathbb{E}_{FT}\left\{ K^{a}\right\} \sim \left(\dfrac{N}{2}\right)^{a} \Big[1 + \dfrac{a (5 + a)}{4 N^2} - \nonumber\\
&- \dfrac{a \left(286 + 183 a - 10 a^2 - 3 a^3\right)}{96 N^4} + \mathcal{O}\left(\dfrac{1}{N^6}\right)\Big] \,.
\end{align}
We notice immediately that already the first subdominant term in this expansion is not correct, since it has a spurious quadratic dependence on $a$. Due to this dependence, \eqref{bigNK} predicts that the variance ${\rm Var}(K) \to 1/8$ as $N \to \infty$, which is evidently incorrect. 
Therefore, albeit \eqref{bigNK} may serve as useful approximation (having the same level of accuracy as an approximate form of $P(x_{min}^{(GUE)} \geq \sqrt{2 N} \xi)$ in \cite{DM}), the only reliable term in it is the first leading term, \eqref{bigNK1}.

\end{document}